# Minimizing the Continuous Diameter when Augmenting a Geometric Tree with a Shortcut[☆],[☆☆]

Jean-Lou De Carufel[a], Carsten Grimm[b,c,∗], Anil Maheshwari[c],
Stefan Schirra[b], Michiel Smid[c]

[a]*School of Electrical Engineering and Computer Science, University of Ottawa*
*800 King Edward Avenue, Ottawa, Ontario, K1N 6N5, Canada*
[b]*Institut für Simulation und Graphik, Otto-von-Guericke-Universität Magdeburg*
*Universitätsplatz 2, D-39106 Magdeburg, Germany*
[c]*School of Computer Science, Carleton University*
*1125 Colonel By Drive, Ottawa, Ontario, K1S 5B6, Canada*

**Abstract**

We augment a tree $T$ with a shortcut $pq$ to minimize the largest distance between any two points along the resulting augmented tree $T + pq$. We study this problem in a continuous and geometric setting where $T$ is a geometric tree in the Euclidean plane, a shortcut is a line segment connecting any two points along the edges of $T$, and we consider all points on $T + pq$ (i.e., vertices and points along edges) when determining the largest distance along $T + pq$. The *continuous diameter* is the largest distance between any two points along edges. We establish that a single shortcut is sufficient to reduce the continuous diameter of a geometric tree $T$ if and only if the intersection of all diametral paths of $T$ is neither a line segment nor a point. We determine an optimal shortcut for a geometric tree with $n$ straight-line edges in $O(n \log n)$ time.

*Keywords:* Geometric Network, Augmentation, Continuous Diameter, Shortcut, Tree
*2010 MSC:* 68U05, 05C85

## 1. Introduction

A *network* is a connected, undirected graph with positive edge weights. A curve is rectifiable if it has a finite length. A *geometric network* is a network that is embedded in the Euclidean plane whose edges are rectifiable curves weighted with their length. We describe our algorithmic results for straight-line edges, even though they extend to more general edges. We define points along the edges of a network as follows. Let $uv \in E$ be an edge in $G$ with weight $w_{uv}$ that connects the vertices $u, v \in V$. For every value $\lambda \in [0, 1]$, we define a point $p$ on edge $uv$ that subdivides $uv$ into two sub-edges $up$ and

---

[☆]This work was partially supported by NSERC and FQRNT.
[☆☆]A preliminary version of this work was presented at the 15th International Symposium on Algorithms and Data Structures (WADS 2017), July 31 to August 2, 2017, St. John's, NL, Canada.
[∗]Corresponding author

$pv$ of weights $w_{up} = \lambda w_{uv}$ and $w_{pv} = (1 - \lambda)w_{uv}$, respectively. We write $p \in uv$ to indicate that $p$ is a point along the edge $uv$, for some $\lambda \in [0, 1]$, and we write $p \in G$ to denote that $p$ is a point along some edge of the network $G$. There is no ambiguity if two edges cross: there are two points along the network that correspond to the crossing in the plane, since points along edges are specified by their relative position to the endpoints of their containing edge, expressed by $\lambda$, and not by coordinates in the plane.

The *network distance* between any two points $p$ and $q$ on a geometric network $G$ is the length of a shortest weighted path from $p$ to $q$ in $G$ and it is denoted by $d_G(p, q)$. The *continuous diameter* of $G$ is the largest network distance between any two points on $G$, and it is denoted by $\operatorname{diam}(G)$, i.e., $\operatorname{diam}(G) = \max_{p,q \in G} d_G(p, q)$. In contrast, for a network with vertex set $V$, the *discrete diameter* is the largest distance between any two vertices, i.e., $\max_{u,v \in V} d_G(u, v)$. A pair $p, q \in G$ is *diametral* when their distance is the continuous diameter, i.e., $\operatorname{diam}(G) = d_G(p, q)$. A point $p \in G$ is a *diametral partner* in $G$ if there exists some point $q \in G$ such that $p, q$ is a diametral pair of $G$. A *diametral path* in $G$ is a shortest weighted path in $G$ that connects a diametral pair of $G$.

We denote the Euclidean distance between two points $p$ and $q$ by $|pq|$. A line segment $pq$ with endpoints $p, q \in G$ is a *shortcut* for $G$. We *augment* a geometric network $G$ with a shortcut $pq$: If they do not exist already, we introduce vertices at $p$ and at $q$, thereby subdividing the edges containing $p$ and $q$. We add the line segment $pq$ as an edge of length $|pq|$ to $G$ without introducing vertices at crossings between $pq$ and other edges. We denote the resulting network by $G + pq$. In this work, we move a shortcut along a network and some of the intermediate shortcuts may coincide with edges or parts of edges. To simplify this discussion, we allow shortcuts $pq$ for a network $G$ with $d_G(p, q) = |pq|$. Instead a local constraint on the definition of a shortcut, we consider a shortcut to be *useful* for $G$ when its addition to $G$ reduces the continuous diameter, i.e., $\operatorname{diam}(G + pq) < \operatorname{diam}(G)$.

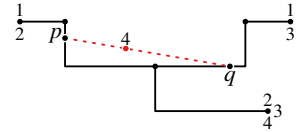

Figure 1: An optimal shortcut $pq$ for a geometric tree $T$ with the diametral pairs of $T + pq$ marked by matching numbers.

Our goal is to locate a shortcut $pq$ for a geometric tree $T$ that minimizes the continuous diameter of the augmented tree $T + pq$, as illustrated in Figure 1. This means we seek two points $p, q \in T$ with $\operatorname{diam}(T + pq) = \min_{r,s \in T} \operatorname{diam}(T + rs)$. We call a shortcut that minimizes the continuous diameter an *optimal shortcut*.

The *backbone* of a tree $T$ is the intersection of all diametral paths of $T$; we denote the backbone of $T$ by $\mathcal{B}$. The *absolute center* of a geometric tree $T$ is the unique point $c \in T$ that minimizes the largest network distance from $c$, i.e., $\max_{q \in T} d_T(c, q) = \min_{p \in T} \max_{q \in T} d_T(p, q)$. We always have $c \in \mathcal{B}$. It takes $O(n)$ time to determine the absolute center—and, thus, the backbone—of a geometric tree with $n$ vertices [1].

*Our Contributions.* We obtain the following structural and algorithmic results:
1. A geometric tree $T$ admits a useful shortcut if and only if its backbone $\mathcal{B}$ is neither a straight-line segment nor a point (i.e., a degenerate line segment).
2. Every geometric tree $T$ has an optimal shortcut $pq$ with both endpoints on the backbone, i.e., $p, q \in \mathcal{B}$, and the absolute center $c$ on the path from $p$ to $q$ in $T$.
3. We develop an algorithm that produces an optimal shortcut for a geometric tree $T$ with $n$ vertices whose edges are straight-line segments in $O(n \log n)$ time.

*1.1. Related Work*

We summarize related work on minimum-diameter network augmentation.

In the *abstract and discrete setting*, the goal is to minimize the discrete diameter of an abstract graph $G = (V, E)$ with positive weights for the edges of $G$ and the edges of its complement graph $\bar{G} = (V, \binom{V}{2} - E)$ by inserting edges of $\bar{G}$ as shortcuts to $G$. If the edges of $G$ and $\bar{G}$ have unit weight, then it is NP-hard to decide whether the diameter can be reduced below some target value $D \geq 2$ by adding at most $k$ shortcuts [2–4]. This problem remains NP-hard when the number of shortcuts is variable, even for trees [2]. Its weighted version falls into the parameterized complexity class W[2]-hard [5, 6]. On the other hand, Oh and Ahn [7] determine an optimal vertex-to-vertex shortcut that minimizes the discrete diameter of an abstract $n$-vertex tree with positive edge weights in $O(n^2 \log^3 n)$ time. Minimum-diameter augmentation has also been studied as a bicriteria optimization in which both the diameter and the number (or cost) of the additional edges are minimized. For instance, Frati et al. [5] summarize the literature regarding the research on bicriteria optimization in minimum-diameter network augmentation.

Große et al. [8] introduce the *geometric and discrete setting* in which the problem is to minimize the discrete diameter of a geometric network by connecting vertices with line segments. Große et al. [8] determine an optimal shortcut for a polygonal path with $n$ vertices in $O(n \log^3 n)$ time using a parametric search technique. Recently, Wang [9] improved this algorithm to $O(n)$ time. Apart from the discrete diameter, the stretch factor, i.e., the largest ratio of the network distance between any two vertices and their Euclidean distance, has also been considered as a target function [10, 11].

In the *geometric and continuous setting* [12], the task is to minimize the continuous diameter of a geometric network by inserting line segments that may connect any two points along the edges. For a polygonal path of length $n$, one can determine an optimal shortcut in $O(n)$ time. For a cycle, one shortcut can never decrease the continuous diameter while two always suffice. For convex cycles with $n$ vertices, one can determine an optimal pair of shortcuts in $O(n)$ time. In the model studied in this work, a crossing of a shortcut with an edge or another shortcut is not a vertex: a path may only enter edges at their endpoints. In the *planar model* [13, 14], every crossing is a vertex of the resulting network, which leads to a different graph structure and, thus, continuous diameter. In the planar model, Yang [14] characterizes optimal shortcuts for a polygonal path. Cáceres et al. [13] determine in polynomial time whether the continuous diameter of a plane geometric network can be reduced with a single shortcut.

*1.2. Structure and Results*

This work is structured as follows. In Section 2, we establish our structural results: We observe that—unlike in the discrete version of this problem—adding a shortcut to a tree might increase the continuous diameter. We characterize the trees that have a useful shortcut, i.e., a shortcut that reduces the continuous diameter, as precisely those trees where the intersection of all diametral paths is neither a straight-line segment nor a point (i.e., a degenerate line segment). The intersection of all diametral paths, called the backbone $\mathcal{B}$ of $T$ plays a key role when locating an optimal shortcut. In the discrete setting, Große et al. [15] show that there exists an optimal shortcut for a tree with both endpoints along the backbone. We prove that this result carries over to the continuous setting and strengthen it: we show that every geometric tree has an optimal shortcut $pq$ with both endpoints along the backbone such that the absolute center $c$ of $T$ lies on the path from $p$ to $q$ in $T$. This yields a restriction of the search space that allows us to find an optimal shortcut for a geometric tree with $n$ straight-line edges in $O(n \log n)$ time. We develop this algorithm in three steps: In Section 4, we examine how the diametral paths in the augmented tree rule out certain directions for the search. In Section 5, we develop a set of rules that inform us how to continuously slide a shortcut along the backbone until we eventually reach an optimal shortcut. In Section 6, we simulate this conceptual continuous algorithm with a discretization that achieves the desired running time.

The structural results hold for geometric tree whose edges are rectifiable curves, i.e., curves that have a well defined length. We describe the algorithmic result for trees with straight-line edges; the techniques carry over to more general types of edges, e.g., algebraic curves of constant degree. Our model of computation is the real RAM.

## 2. Usefulness

We say a shortcut $pq$ is *useful* for $T$ when $\text{diam}(T + pq) < \text{diam}(T)$, we say $pq$ is *indifferent* for $T$ when $\text{diam}(T + pq) = \text{diam}(T)$, and we say $pq$ is *useless* for $T$ when $\text{diam}(T + pq) > \text{diam}(T)$. In the discrete setting, every shortcut is useful or indifferent, as the discrete diameter only considers vertices of $T$. In the continuous setting, a shortcut may be useless for $T$, since the points on the shortcut $pq$ matter as well, as in Figure 2.

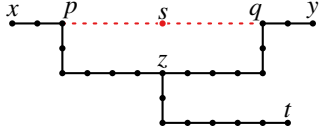
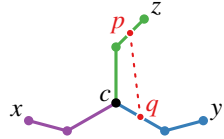

Figure 2: A geometric tree $T$ with a shortcut $pq$ that is useless for $T$. Each edge has a unit weight and $|pq| = 8$. The continuous diameter increases when we augment $T$ with $pq$, since $\text{diam}(T) = d_T(x, y) = 16$ and $\text{diam}(T + pq) = d_{T+pq}(s, t) = 17$.

Figure 3: A geometric tree $T$ whose continuous diameter cannot be reduced with a single shortcut. Each pair of the three leaves $x$, $y$, and $z$ is diametral, as $d_T(x, c) = d_T(y, c) = d_T(z, c)$. The shortcut $pq$ is not useful for $T$, as it is indifferent for $\{x, y\}$.

For two points $u, v \in T$, a shortcut $pq$ is *useful* for the unordered pair $\{u, v\}$ when $d_{T+pq}(u, v) < d_T(u, v)$, and $pq$ is *indifferent* for $\{u, v\}$ when $d_{T+pq}(u, v) = d_T(u, v)$.

We say a shortcut $pq$ is *useful* for the ordered pair $(u, v)$ when $d_T(u, p) + |pq| + d_T(q, v) < d_T(u, v)$, we say $pq$ is *indifferent* for $(u, v)$ when $d_T(u, p) + |pq| + d_T(q, v) = d_T(u, v)$, and we say that $pq$ is *useless* for $(u, v)$ when $d_T(u, p) + |pq| + d_T(q, v) > d_T(u, v)$. If $pq$ is useful for $(u, v)$ then the shortest path from $u$ to $v$ in $T + pq$ travels from $u$ to $p$ in $T$, then along the shortcut to $q$, and then from $q$ to $v$ in $T$. The order matters for ordered pairs: if $pq$ is useful for the ordered pair $(u, v)$, then $pq$ is useless for the ordered pair $(v, u)$, and $qp$ is useless for the ordered pair $(u, v)$. We only distinguish $pq$ and $qp$ for ordered pairs; there is no difference between $pq$ and $qp$ and, thus, $T + pq = T + qp$.

For some trees, we cannot reduce the continuous diameter with a single shortcut. For instance, every shortcut $pq$ for the tree $T$ in Figure 3 is either useless or indifferent for $T$, since $pq$ would be indifferent for at least one of the diametral pairs $\{x, y\}$, $\{x, z\}$, or $\{y, z\}$ of $T$. In Figure 3, the intersection of the diametral paths of $T$ consists of a single vertex. We argue that a tree with this property cannot have a useful shortcut.

**Lemma 1** *If the backbone of a geometric tree $T$ consists only of the absolute center of $T$, then there does not exists a useful shortcut for $T$.*

PROOF. Suppose $T$ is a geometric tree whose backbone $\mathcal{B}$ consists only of the absolute center $c$ of $T$. Then, $c$ must be a vertex of $T$, since $\mathcal{B}$ would contain the entire edge containing $c$, otherwise. Let $x, y$ be a diametral pair of $T$. Then we have $\frac{1}{2} \operatorname{diam}(T) = d_T(x, c) = d_T(y, c)$, since $c$ is the midpoint of the path from $x$ to $y$ in $T$. Since the backbone only consists of $c$, there must be at least one other leaf $z$ with $\frac{1}{2} \operatorname{diam}(T) = d_T(z, c)$ such that the paths from $z$ to $x$ and from $z$ to $y$ pass through $c$. This means that $x$, $y$, and $z$ lie in three different sub-trees $T_X$, $T_Y$, and $T_Z$ attached to $c$.

Assume, for the sake of a contradiction, that there exists a shortcut $pq$ that is useful for $T$. Then $p \neq q$, since $pq$ would be indifferent for $T$, otherwise. Hence, we have $p \neq c$ or $q \neq c$, and one of the sub-trees $T_X$, $T_Y$, or $T_Z$ contains neither $p$ nor $q$. Without loss of generality, suppose $p, q \notin T_X$. By our assumption, $pq$ must be useful for every diametral pair of $T$ including $\{x, y\}$. Without loss of generality, let $pq$ be useful for $(x, y)$, i.e., $d_{T+pq}(x, y) = d_T(x, p) + |pq| + d_T(q, y) < d_T(x, y)$. Otherwise, we swap $p$ and $q$. This implies that $q$ lies in $T_Y$, since otherwise the shortest path in $T + pq$ from $x$ to $y$ would contain $c$ twice: once along the sub-path from $x$ to $p$ and once along the sub-path from $q$ to $y$. Since $pq$ is useful for $T$ by assumption, $pq$ must be useful for the unordered pair $\{x, z\}$ and, thus, $pq$ must be useful for either $(x, z)$ or $(z, x)$. The shortcut $pq$ cannot be useful for the ordered pair $(x, z)$, since the path from $q$ to $z$ contains $c$. Therefore, $pq$ is useful for $(z, x)$ and, thus, $p$ lies in $T_Z$, since the path from $x$ to $q$ in $T + pq$ contains $c$.

In summary, the shortcut $pq$ is useful for $(x, y)$ and for $(z, x)$ with $p \in T_Z$ and $q \in T_Y$. This is impossible: Along the shortest path from $x$ to $y$ in $T + pq$ the sub-path from $c$ to $p$ *does not* contain $pq$, i.e., $d_{T+pq}(c, p) = d_T(c, p)$. On the other hand, along the shortest path from $x$ to $z$ in $T + pq$, the sub-path from $c$ to $p$ *does* contain $pq$, i.e., $d_{T+pq}(c, p) < d_T(c, p)$. Therefore, there does not exist a single useful shortcut for $T$. □

Let $T$ be a geometric tree whose backbone $\mathcal{B}$ does not consist of a single vertex, as illustrated in Figure 4. In this case, the backbone $\mathcal{B}$ is a path with endpoints $a$ and $b$ such that $a \neq b$. We call the sub-trees that are attached to $\mathcal{B}$ the $\mathcal{B}$-*sub-trees* of $T$. The *root* of a $\mathcal{B}$-sub-tree $S$ is the vertex $r$ connecting $S$ and $\mathcal{B}$. Let $X$ be the $\mathcal{B}$-sub-tree with root $a$, and let $Y$ be the $\mathcal{B}$-sub-tree with root $b$. We refer to $X$ and $Y$ as the *primary $\mathcal{B}$-sub-trees*, because every diametral pair of $T$ consists of a leaf $x$ in $X$ and of a leaf $y$ in $Y$. The other $\mathcal{B}$-sub-trees $S_1, S_2, \ldots, S_k$ with roots $r_1$, $r_2, \ldots, r_k$, respectively, are the *secondary $\mathcal{B}$-sub-trees* of $T$. Each primary $\mathcal{B}$-sub-trees may consist only of an endpoint of the backbone, as in Figure 4.

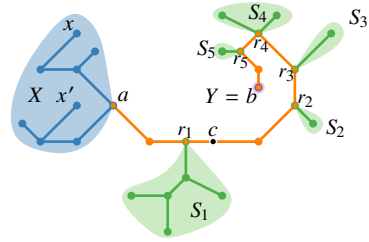

Figure 4: A geometric tree $T$ with its backbone $\mathcal{B}$ (orange), primary $\mathcal{B}$-sub-trees, $X$ (blue) and $Y$ (purple), and secondary $\mathcal{B}$-sub-trees (green). The primary $\mathcal{B}$-sub-tree $Y$ coincides with the endpoint $b$ of $\mathcal{B}$.

**Theorem 2** *For geometric tree $T$ with backbone $\mathcal{B}$, there exist two points $p, q \in T$ such that $pq$ is a useful shortcut for $T$ if and only if $\mathcal{B}$ is not a line segment (or a point).*

PROOF. Let $T$ be a geometric tree and let $pq$ be a useful shortcut for $T$. We show that the backbone $\mathcal{B}$ of $T$ is neither a line segment nor a point (i.e., a degenerate line segment). Since $T$ possesses a useful shortcut, the backbone $\mathcal{B}$ of $T$ cannot be a point, due to Lemma 1. Let $a$ and $b$ be the endpoints of the path $\mathcal{B}$. Let $X$ and $Y$ be the primary $\mathcal{B}$-sub-trees attached to $a$ and $b$, respectively. Let $x, y$ be a diametral pair of $T$ with $x \in X$ and $y \in Y$. Since $pq$ is useful for $T$, the shortcut $pq$ must be useful for $\{x, y\}$. Without loss of generality, $pq$ is useful for $(x, y)$. Otherwise, we swap $p$ and $q$.

We argue that the backbone $\mathcal{B}$ differs from the line segment $ab$ that connects its endpoints by showing that $pq$ is useful for $(a, b)$, since this implies $|ab| < d_T(a, b)$. We locate a diametral pair $\hat{x}, \hat{y}$ of $T$ with $\hat{x} \in X$ and $\hat{y} \in Y$ such that $a$ lies on the path from $\hat{x}$ to $p$, and $b$ lies on the path from $q$ to $\hat{y}$, and $pq$ is useful for $(\hat{x}, \hat{y})$. The existence of a diametral pair $\hat{x}, \hat{y}$ of $T$ with these properties implies that $pq$ is useful for $(a, b)$, i.e., $d_T(a, p) + |pq| + d_T(q, b) < d_T(a, b)$, because then

$$d_T(\hat{x}, a) + d_T(a, p) + |pq| + d_T(q, b) + d_T(b, \hat{y})$$
$$= d_T(\hat{x}, p) + |pq| + d_T(q, \hat{y}) < d_T(\hat{x}, \hat{y}) = d_T(\hat{x}, a) + d_T(a, b) + d_T(b, \hat{y}) \ .$$

We start with $x, y$ as a candidate for $\hat{x}, \hat{y}$. If $a$ lies lies on the path from $x$ to $p$, then we let $\hat{x} = x$. If $a$ does not lie on the path from $x$ to $p$, then $p \in X$ with $p \neq a$. Since $a$ is an endpoint of the intersection of the diametral paths in $T$, there is some leaf $x'$ of $X$ such that $x', y$ is diametral in $T$ and $a$ lies on the path from $x'$ to $p$. Let $\hat{x} = x'$. We argue that $pq$ is useful for $(x', y)$. Since $pq$ is useful for $T$, it must be useful for $\{x', y\}$. The shortcut $pq$ is useful for $(x, y)$ and, therefore, useful for $(p, y)$. Thus, the shortest path connecting $p$ and $y$ in $T + pq$ contains $q$. If $pq$ was useful for $(y, x')$, then the shortest path connecting $x'$ and $y$ in $T + pq$ would travel from $x'$ to $q$, then via the shortcut $qp$ to $p$, and then backwards via $q$ to $y$. This is impossible and $pq$ must be useful for $(x', y)$. Likewise, we find $\hat{y}$ where $pq$ is useful for $(\hat{x}, \hat{y})$ and $b$ lies on the path from $q$ to $\hat{y}$.

Therefore, $pq$ is useful for $(a, b)$. By the triangle inequality, $|ab| \leq d_T(a, p) + |pq| + d_T(q, b)$ and, thus, $|ab| < d_T(a, b)$. The backbone is strictly longer than the line segment connecting its endpoints and, thus, the backbone cannot be a line segment itself.

Conversely, suppose $T$ is a geometric tree whose backbone $\mathcal{B}$ is neither a line segment nor a single point, as in Figure 5. Thus, the backbone $\mathcal{B}$ of $T$ is a path with endpoints $a$ and $b$ such that $|ab| < d_T(a, b)$. Let $X$ and $Y$ be the primary $\mathcal{B}$-sub-trees of $T$ attached to $a$ and to $b$, respectively, and let $S_1, S_2, \ldots, S_k$ be the secondary $\mathcal{B}$-sub-trees of $T$.

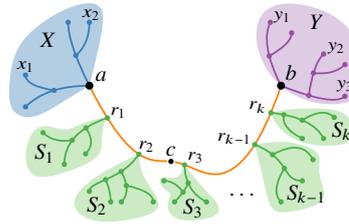

Figure 5: A geometric tree whose backbone is a path from $a$ to $b$ that is not a line segment.

Let $s, t$ be a diametral pair of $T + ab$. We distinguish three cases: we have $s, t \in T$ and $s, t$ is a diametral in $T$, or we have $s, t \in T$ and $s, t$ is not diametral in $T$, or we have $s \notin T$ or $t \notin T$. In the first two cases, we show that $ab$ is useful for $T$, and in the third case we construct a useful shortcut $pq$ for $T$ with $p, q \in \mathcal{B}$ if $ab$ is not useful for $T$.

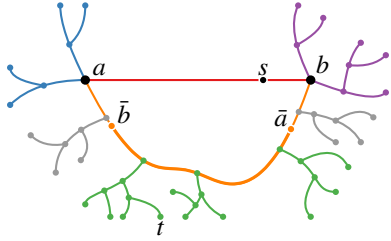 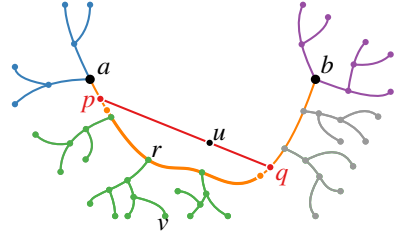

Figure 6: A sketch of a geometric tree $T$ whose backbone is not a line segment connecting its endpoints $a$ and $b$. A diametral pair $s, t$ of $T + ab$ with $s \notin T$. The diametral partner $t$ lies in a $\mathcal{B}$-sub-tree attached to the path from $\bar{a}$ to $\bar{b}$ in $T$, where $\bar{a}$ and $\bar{b}$ are the respective farthest points from $a$ and from $b$ along the simple cycle $C(a, b)$ in $T + ab$.

Figure 7: A sketch of a geometric tree whose backbone is not a line segment connecting its endpoints $a$ and $b$. We moved the shortcut to a position $pq$, since the shortcut $ab$ increased the diameter. The points $p$ and $q$ are chosen such that the simple cycle $C(p, q)$ in $T + pq$ is too small to contain the depicted diametral pair $u, v$ with $u \notin T$.

1. Suppose $s, t \in T$ such that $s, t$ is diametral in $T$, i.e., $d_T(s, t) = \text{diam}(T)$.
   Then $s$ and $t$ lie in different primary $\mathcal{B}$-sub-trees of $T$. Without loss of generality, we assume that $s \in X$ and $t \in Y$. Otherwise, we swap $s$ and $t$.
   The shortcut $ab$ is useful for $(s, t)$, i.e., $d_T(s, a) + |ab| + d_T(b, y) < d_T(s, t)$, since

   $$d_T(s, a) + |ab| + d_T(b, y) < d_T(s, a) + d_T(a, b) + d_T(b, y) = d_T(s, t) \ .$$

   Therefore, $d_{T+ab}(s, t) < d_T(s, t)$ and, thus, the shortcut $ab$ is useful for $T$, since

   $$\text{diam}(T + ab) = d_{T+ab}(s, t) < d_T(s, t) = \text{diam}(T) \ .$$

2. Suppose $s, t \in T$ such that $s, t$ is not diametral in $T$, i.e., $d_T(s, t) < \text{diam}(T)$.
   Then the shortcut $ab$ is useful for $T$, since

   $$\text{diam}(T + ab) = d_{T+ab}(s, t) \leq d_T(s, t) < \text{diam}(T) \ .$$

3. Suppose $s \notin T$ or $t \notin T$. Without loss of generality, let $s \notin T$.
   Then $s \in ab$ with $a \neq s \neq b$. Let $C(a, b)$ be the simple cycle in $T + ab$. We distinguish two cases depending on whether $t$ lies on $C(a, b)$ or not.
   (a) Suppose $t \in C(a, b)$. Then we have $d_{T+ab}(s, t) < d_T(a, b)$, because

   $$d_{T+ab}(s, t) \leq \frac{|ab| + d_T(a, b)}{2} < \frac{d_T(a, b) + d_T(a, b)}{2} = d_T(a, b) \ .$$

   Thus, $ab$ is useful for $T$, as $\text{diam}(T+ab) = d_{T+ab}(s, t) < d_T(a, b) \leq \text{diam}(T)$.
   (b) Suppose $t \notin C(a, b)$, as illustrated in Figure 6.
   We argue that $t$ is a leaf of a secondary $\mathcal{B}$-sub-tree. Let $\bar{a}$ be the farthest point on $C(a, b)$ from $a$ with respect to $T + ab$. The point $\bar{a}$ lies in $T$, because

   $$|ab| = \frac{|ab| + |ab|}{2} < \frac{|ab| + d_T(a, b)}{2} = d_{T+ab}(a, \bar{a}) \ .$$

   Therefore, $t$ cannot lie in $X$, because $\bar{a}$ is the farthest point on $C(a, b)$ from any point in $X$, and $s \neq \bar{a}$, since $s \notin T$ and $\bar{a} \in T$. Likewise, $t \notin Y$. In summary, $s$ lies on the shortcut $ab$ and $t$ is a leaf of a secondary $\mathcal{B}$-sub-tree.

Let $\delta$ be the largest diameter of the secondary $\mathcal{B}$-sub-trees $S_1, \ldots, S_k$, i.e.,

$$\delta := \max_{i=1}^{k} \text{diam}(S_i) \ ,$$

and let $\epsilon = \text{diam}(T) - \delta$. We have $\epsilon > 0$, as none of the secondary $\mathcal{B}$-sub-trees contains diametral pairs of $T$, i.e., $\text{diam}(S_i) < \text{diam}(T)$ for all $i = 1, 2, \ldots, k$. Since $|ab| < d_T(a, b)$, there exist $p, q \in \mathcal{B}$ with $|pq| < d_T(p, q)$ and $|pq| + d_T(p, q) < 2\epsilon$. We argue that $pq$ is useful for $T$. Let $u, v$ be a diametral pair of $T + pq$ and let $C(p, q)$ be the simple cycle in $T + pq$. Analogously to diametral pairs of $T + ab$, we argue that $pq$ is useful for $T$ when $u, v \in T$ (Case 1 and 2) and when $u \in pq$ with $p \neq u \neq q$ and $v \in C(p, q)$ (Case 3a). It remains to show that $pq$ is useful for $T$ when $u \in pq$ with $p \neq u \neq q$ and when $v$ is a leaf of a secondary $\mathcal{B}$-sub-tree that is attached to an interior vertex $r$ of the path from $p$ to $q$ in $T$ (Case 3b), as illustrated in Figure 7. By definition, $d_T(v, r) \leq \delta$. By choice of $p$ and $q$,

$$d_{T+pq}(r, u) \leq \frac{|C(p, q)|}{2} = \frac{|pq| + d_T(p, q)}{2} < \frac{2 \cdot \epsilon}{2} = \epsilon \ .$$

Therefore, the shortcut $pq$ is useful for $T$, since

$$\begin{aligned}
\text{diam}(T + pq) &= d_{T+pq}(u, v) &&(u, v \text{ is diametral in } T + pq) \\
&= d_T(v, r) + d_{T+pq}(r, u) \\
&&&(r \text{ is on any path from } v \text{ to } u \text{ in } T + pq) \\
&< d_T(v, r) + \epsilon &&(d_{T+pq}(r, u) < \epsilon) \\
&\leq \delta + \epsilon &&(d_T(v, r) \leq \delta) \\
&= \delta + \text{diam}(T) - \delta &&(\epsilon = \text{diam}(T) - \delta) \\
&= \text{diam}(T) \ .
\end{aligned}$$

Every diametral pair $s, t$ of $T + ab$ falls into one of the above cases and, in each case, we either argued that $ab$ is useful for $T$ or found a useful shortcut for $T$ when $ab$ was not useful for $T$. Therefore, $T$ possesses a useful shortcut when $\mathcal{B}$ is not a line segment. □

## 3. Optimal Shortcuts

Consider a geometric tree $T$ whose backbone is a path from $a$ to $b$. This path contains the absolute center $c$. We prove that there is an optimal shortcut $pq$ for $T$ such that $p$ lies on the path from $a$ to $c$ and $q$ lies on the path from $c$ to $b$. This holds when $\mathcal{B}$ consists only of $c$, since then $T$ has no useful shortcuts and the degenerate shortcut $cc$ is optimal. For the other cases, we establish our claim by proving the following statements.

1. If $pq$ is a useful shortcut for a geometric tree $T$, then every $\mathcal{B}$-sub-tree $S$ of $T$ contains at most one endpoint of $pq$, i.e., we have $p \notin S$ or $q \notin S$.
2. If an endpoint $p$ of a useful, optimal shortcut $pq$ for $T$ lies in a $\mathcal{B}$-sub-tree $S$ with root $r$, then the shortcut $rq$ is also optimal for $T$—regardless of the position of $q$.
3. There exists an optimal shortcut $pq$ for $T$ with $p, q \in \mathcal{B}$.

4. If $pq$ is an optimal shortcut for $T$ with $p, q \in \mathcal{B}$ and the path from $p$ to $q$ along $\mathcal{B}$ does not contain $c$, then at least one of $pc$ or $cq$ is an optimal shortcut for $T$.

The last statement implies our claim, since $c$ lies on the path from $c$ to $b$ and the path from $c$ to $a$. We prove Statements 1 through 4 in Lemmas 4 to 6 and Theorem 7.

The idea for restricting the search for an optimal shortcut along the backbone stems from Große et al. [15] who establish the following for the discrete setting [15, Lemma 6]. We generalize their result to the continuous setting and incorporate the absolute center.

**Theorem 3 (Discrete Backbone Theorem by Große et al. [15])** *Let $T$ be a geometric tree with vertex set $V$ and backbone $\mathcal{B}$. There exists a pair of vertices $p, q \in \mathcal{B} \cap V$ such that $pq$ minimizes the discrete diameter of the augmented tree $T + pq$ among all possible discrete shortcuts for $T$, i.e.,*

$$\max_{u,v \in V} d_{T+pq}(u, v) = \min_{r,s \in V} \max_{u,v \in V} d_{T+rs}(u, v) \ .$$

**Lemma 4** *Let $T$ be a geometric tree, let $pq$ be a shortcut for $T$, and let $S$ be a $\mathcal{B}$-sub-tree of $T$. If the shortcut $pq$ is useful for $T$, then $p \notin S$ or $q \notin S$.*

PROOF. Assume, for a contradiction, that there is a geometric tree $T$, with a $\mathcal{B}$-sub-tree $S$ with root $r$, such that there is a shortcut $pq$ for $T$ with $p, q \in S$ that is useful for $T$.

Suppose the backbone $\mathcal{B}$ of $T$ is a path that connects the vertices $a$ and $b$. Let $X$ and $Y$ be the primary $\mathcal{B}$-sub-trees with roots $a$ and $b$, respectively. Furthermore, let $x, y$ be a diametral pair of $T$ with $x \in X$ and $y \in Y$. Since $pq$ is useful for $T$, we have $p \neq q$ and $pq$ is useful for $(x, y)$ or for $(y, x)$. Without loss of generality, let $pq$ be useful for $(x, y)$. Otherwise, we swap $x, y$ and $X, Y$. We distinguish whether the root $r$ of $S$ lies on the path from $x$ to $p$ in $T$ or not and we derive a contradiction in both of these cases.

Suppose $r$ *does not* lie on the path from $x$ to $p$. Then $x$, $p$, and $q$ lie in the same $\mathcal{B}$-sub-tree, i.e., $S = X$ and $r = a$, since $x \in X$, as in Figure 8. Since $y$ lies in the other primary $\mathcal{B}$-sub-tree, we have $y \notin S$ and, thus, $r$ lies on the path from $q$ to $y$.

Since $a$ is an endpoint of the backbone, i.e., the intersection of all diametral paths in $T$, there is a leaf $x'$ of $X$ such that $x', y$ is diametral in $T$ and $r$ is on the path from $x'$ to $p$. The shortcut $pq$ cannot be useful for $(x', y)$, since $r$ lies on the paths from $x'$ to $p$ and from $q$ to $y$. Thus, for $pq$ to be useful for $T$, the shortcut $pq$ must be useful for $(y, x')$. The shortest path from $y$ to $x'$ in $T + pq$ contains the path from $r$ to $p$ in $T$, i.e., $d_T(r, p) < d_T(r, q) + |qp|$.

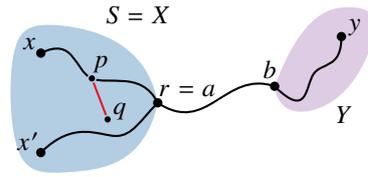

Figure 8: A shortcut $pq$ for $T$ that has both endpoints in the same $\mathcal{B}$-sub-tree $S$ with root $r$ and where $r$ does not lie on the path from $x$ to $p$, which implies $S = X$.

On the other hand, the shortest path from $y$ to $x$ in $T + pq$ travels from $r$ via the shortcut to $p$ and, thus, $d_T(r, q) + |qp| < d_T(r, p)$. This is impossible.

Suppose $r$ lies on the path from $x$ to $p$. Then $r$ *does not* lie on the path from $y$ to $q$ and, thus, $y$, $q$, and $p$ lie in the same $\mathcal{B}$-sub-tree, meaning $S = Y$ and $r = b$ and $x \notin S$. This situation is symmetric to the previous case and, therefore, impossible as well.

Therefore, if the shortcut $pq$ is useful for $T$, then we have $p \notin S$ or $q \notin S$. □

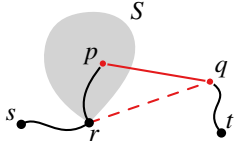

Figure 9: A sketch of the shortest path in $T + pq$ from $s$ via $pq$ to $t$, where $p$ lies in some $\mathcal{B}$-sub-tree $S$ whose root $r$ lies on the path from $s$ to $p$ in $T$. We have $q \notin S$ by Lemma 4, and, thus, $t \notin S$.

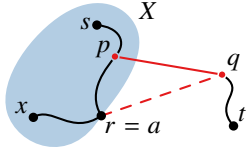

(a) The case $S = X$.

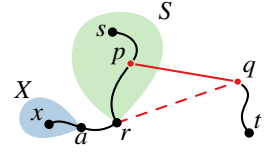

(b) The case $S \neq X$.

Figure 10: A sketch of the shortest path in $T + pq$ connecting $s$ and $t$ via $pq$ where $p$ lies in some $\mathcal{B}$-sub-tree $S$ and $p$ lies on the path from $s$ to the root $r$ of $S$. There is a diametral partner $x$ in $T$ such that the path from $x$ to $p$ passes through $r$. This holds when $S$ is (a) primary or (b) secondary.

**Lemma 5** *Let $T$ be a geometric tree and let $S$ be a $\mathcal{B}$-sub-tree of $T$ with root $r$. If $pq$ is a useful shortcut for $T$ with $p \in S$, then $\mathrm{diam}(T + rq) \leq \mathrm{diam}(T + pq)$.*

The proof of Lemma 5 follows the proof of the discrete backbone theorem [15, Lemma 6]. The first two cases are due to Große et al. [15]; they are provided for self-containment. We add the third case to generalize the result to the continuous setting.

PROOF (PROOF OF LEMMA 5). Let $T$ be a geometric tree, let $S$ be a $\mathcal{B}$-sub-tree of $T$ with root $r$, let $pq$ be a useful shortcut for $T$ with $p \in S$, and let $s, t$ be a diametral pair of $T + rq$. Either we have $s, t \in T$ and $pq$ is indifferent for $\{s, t\}$, or we have $s, t \in T$ and $pq$ is useful for $\{s, t\}$, or we have $s \notin T$ or $t \notin T$. In each case, we argue that $d_{T+rq}(s, t) \leq d_{T+pq}(s', t')$ for some $s', t' \in T + pq$ and, thus, $\mathrm{diam}(T + rq) \leq \mathrm{diam}(T + pq)$.

1. Suppose $s, t \in T$ and $pq$ is indifferent for $\{s, t\}$, i.e., $d_T(s, t) = d_{T+pq}(s, t)$.
   Then we have $\mathrm{diam}(T + rq) \leq \mathrm{diam}(T + pq)$, because
   $$\mathrm{diam}(T + rq) = d_{T+rq}(s, t) \leq d_T(s, t) = d_{T+pq}(s, t) \leq \mathrm{diam}(T + pq) \ .$$

2. Suppose $s, t \in T$ and $pq$ is useful for $\{s, t\}$, i.e., $d_T(s, t) > d_{T+pq}(s, t)$.
   Then $pq$ is useful for $(s, t)$ or $(t, s)$. Without loss of generality, $pq$ is useful for $(s, t)$, i.e., $d_T(s, p) + |pq| + d_T(q, t) = d_{T+pq}(s, t) < d_T(s, t)$. Otherwise, we swap $s$ and $t$.
   We distinguish whether the path from $s$ to $p$ in $T$ contains $r$ or not.
   
   (a) Suppose the path from $s$ to $p$ in $T$ contains $r$, as shown in Figure 9.
       Then we have $\mathrm{diam}(T + rq) \leq \mathrm{diam}(T + pq)$, because
       $$\begin{aligned}
       \mathrm{diam}(T + rq) &= d_{T+rq}(s, t) & (s, t \text{ is diametral in } T + rq) \\
       &\leq d_T(s, r) + |rq| + d_T(q, t) & (\text{triangle inequality for } d_{T+rq}) \\
       &\leq d_T(s, r) + d_T(r, p) + |pq| + d_T(q, t) & (\text{triangle inequality } |\cdot|) \\
       &= d_T(s, p) + |pq| + d_T(q, t) & (r \text{ is on the } s\text{-}t\text{-path in } T) \\
       &= d_{T+pq}(s, t) & (pq \text{ is useful for } (s, t)) \\
       &\leq \mathrm{diam}(T + pq) \ .
       \end{aligned}$$

(b) Suppose the path from $s$ to $p$ in $T$ *does not* contain $r$, as shown in Figure 10. Let $a$ and $b$ be the endpoints of the backbone $\mathcal{B}$ of $T$, and let $X$ and $Y$ be the primary $\mathcal{B}$-sub-trees with roots $a$ and $b$, respectively. We find a diametral pair $x, y$ of $T$ such that $x \in X$, $y \in Y$, and $r$ lies on the path from $x$ to $p$. Every path in $T + pq$ that connects a diametral pair of $T$ contains $pq$, since $pq$ is a useful shortcut for $T$. If $S = X$, then there is a diametral pair $x, y$ of $T$ with $x \in X$ and $y \in Y$ such that the path from $x$ to $p$ contains $r = a$, as in Figure 10a. Otherwise, $a$ would not be the endpoint of the backbone. If $S \neq X$, then $r$ lies on the path from $x$ to $p$ for every $x \in X$, since $p \in S$, as in Figure 10b. The shortcut $pq$ is useful for $(x, y)$ and, thus, for $(r, q)$, since $r$ lies on the path from $x$ to $p$. Therefore, no shortest path in $T + pq$ may contain the path connecting $r$ and $q$. This includes the path from $s$ to $t$. Hence, $t \notin S$ and $d_{T+rq}(s, t) = d_T(s, r) + d_{T+rq}(r, t)$. By Lemma 4, $p \in S$ implies $q \notin S$. Since the path from $x$ to $y$ in $T + pq$ cannot contain $r$ two times, we have $y \notin S$ and the shortest path in $T + rq$ from $x$ to $t$ contains $r$, i.e., $d_{T+rq}(x, t) = d_T(x, r) + d_{T+rq}(r, t)$. We have $d_T(s, r) \leq d_T(x, r)$, as $x, y$ is a diametral path of $T$ and $r$ lies on the path from $x$ to $y$ and on the path from $s$ to $y$. This implies that $x, t$ is a diametral pair of $T + rq$, since

$$\begin{aligned}
\text{diam}(T + rq) &= d_{T+rq}(s, t) & (s, t \text{ is diametral in } T + rq) \\
&= d_T(s, r) + d_{T+rq}(r, t) & (s \in S \text{ and } t \notin S) \\
&\leq d_T(x, r) + d_{T+rq}(r, t) & (d_T(s, r) \leq d_T(x, r)) \\
&= d_{T+rq}(x, t) & (r \text{ is on any } x\text{-}t\text{-path in } T + rq) \\
&\leq \text{diam}(T + rq) \ .
\end{aligned}$$

More precisely, $x, t$ is a diametral pair of $T + rq$ such that the path from $x$ to $p$ contains $r$. With the argument from the previous case, we obtain

$$\text{diam}(T + rq) = d_{T+rq}(s, t) = d_{T+rq}(x, t) \leq \text{diam}(T + pq) \ .$$

3. Suppose $s \notin T$ or $t \notin T$. Without loss of generality, let $s \notin T$. Otherwise, we swap $s$ and $t$. Then we have $s \in rq$ with $r \neq s \neq q$, as illustrated in Figure 11.
Let $C(r, q)$ be the simple cycle in $T + rq$. The path from $s$ to $t$ leaves $C(r, q)$ at the point $\bar{s} \in C(r, q)$ with $d_{T+rq}(s, \bar{s}) = \frac{1}{2}(d_T(r, q) + |rq|)$. Note that $r \neq \bar{s} \neq q$, since $r \neq s \neq q$ and $|rq| \leq d_T(r, q)$. By Lemma 4, $p \in S$ implies $q \notin S$.

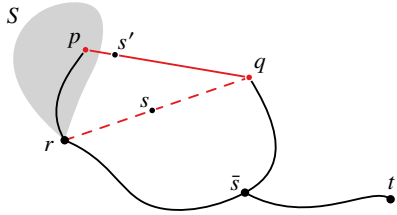

Figure 11: A diametral pair $s, t$ of $T + rq$ with $s \in rq$. We move $r \in \mathcal{B}$ to a point $p$ in a $\mathcal{B}$-sub-tree $S$ such that $pq$ is useful for $T$. Then $q, t \notin S$ and the diameter increases.

This means the path connecting $r$ and $q$ in $T$ lies outside of $S$ and, thus, $\bar{s} \notin S$ and $t \notin S$. The cycle $C(p, q)$ is formed by $pq$ and the path from $p$ to $q$ in $T$. Since $r$ lies on the path from $p$ to $q$, we know that $\bar{s} \in C(p, q)$. Let $s'$ be the farthest point from $\bar{s}$ on $C(p, q)$. Then we have $\mathrm{diam}(T + rq) \leq \mathrm{diam}(T + pq)$, because

$$
\begin{aligned}
\mathrm{diam}(T + rq) &= d_{T+rq}(s, t) & \text{($s, t$ is diametral in $T + rq$)} \\
&= d_{T+rq}(s, \bar{s}) + d_T(\bar{s}, t) & \text{($\bar{s}$ is on any $t$-$s$-path in $T + rq$)} \\
&= \frac{d_T(r, q) + |rq|}{2} + d_T(\bar{s}, t) & \text{($s$ and $\bar{s}$ are antipodals on $C(r, q)$)} \\
&\leq \frac{d_T(r, q) + d_T(r, p) + |pq|}{2} + d_T(\bar{s}, t) & \text{(triangle inequality)} \\
&= \frac{d_T(p, q) + |pq|}{2} + d_T(\bar{s}, t) & \text{($r$ is on the $p$-$q$-path in $T$)} \\
&= d_{T+pq}(s', \bar{s}) + d_T(\bar{s}, t) & \text{($s$ and $\bar{s}'$ are antipodals on $C(p, q)$)} \\
&= d_{T+pq}(s', t) & \text{($\bar{s}$ is on any $t$-$s'$-path in $T + pq$)} \\
&\leq \mathrm{diam}(T + pq) \ .
\end{aligned}
$$

Thus, if $pq$ is a useful shortcut for $T$ with $p \in S$, then $\mathrm{diam}(T + rq) \leq \mathrm{diam}(T + pq)$. □

**Lemma 6 (Continuous Backbone Lemma)** *Let $T$ be a geometric tree $T$ with backbone $\mathcal{B}$. There exist $p, q \in \mathcal{B}$ such that $pq$ is an optimal shortcut for $T$.*

PROOF. Let $pq$ be an optimal shortcut for a geometric tree $T$. We assume that $pq$ is useful for $T$, since otherwise the shortcut $cc$ is indifferent for $T$ with $c \in \mathcal{B}$.

If $p, q \in \mathcal{B}$, we are done. Suppose $p \notin \mathcal{B}$, i.e., $p$ lies in some $\mathcal{B}$-sub-tree $S$ with root $r \in \mathcal{B}$. According to Lemma 5 this implies $\mathrm{diam}(T + rq) \leq \mathrm{diam}(T + pq)$. This means that $rq$ is also an optimal shortcut for $T$. If $q \in \mathcal{B}$, then the claim follows. Otherwise, $q \notin \mathcal{B}$, i.e., $q$ lies in some $\mathcal{B}$-sub-tree $S'$ with root $r' \in \mathcal{B}$. Since $pq$ is useful for $T$ and $p \in S$, Lemma 4 implies that $q \notin S$ and, thus, $S \neq S'$ and $r \neq r'$. By Lemma 5, this implies $\mathrm{diam}(T + rr') \leq \mathrm{diam}(T + rq) \leq \mathrm{diam}(T + pq)$, and, thus, $rr'$ is an optimal shortcut for $T$ with both endpoints along the backbone of $T$. □

**Theorem 7** *Let $T$ be a geometric tree with backbone $\mathcal{B}$ and absolute center $c$. There exist an optimal shortcut $pq$ for $T$ with $p, q \in \mathcal{B}$ and $c$ on the path from $p$ to $q$ in $T$.*

PROOF. Let $T$ be a geometric tree with backbone $\mathcal{B}$ and absolute center $c$. By Lemma 6, there exists an optimal shortcut $pq$ for $T$ with $p, q \in \mathcal{B}$. If $pq$ is indifferent for $T$, then the degenerate shortcut $cc$ satisfies the claim. Thus, we assume that $pq$ is useful for $T$.

Since there exists a useful shortcut for $T$, the backbone $\mathcal{B}$ of $T$ is a path with endpoints $a$ and $b$ that contains $c$ with $a \neq b$. Suppose $c$ does not lie on the path from $p$ to $q$ along $\mathcal{B}$. Without loss of generality, we assume that $p$ and $q$ lie on the path from $a$ to $c$ with $p \neq c \neq q$. Otherwise, we swap $a$ and $b$. Without loss of generality, the path from $p$ to $c$ along $\mathcal{B}$ contains $q$, i.e., $d_T(q, c) \leq d_T(p, c)$. Otherwise, we swap $p$ and $q$.

We argue that $pc$ is at least as good as $pq$, i.e., $\mathrm{diam}(T + pc) \leq \mathrm{diam}(T + pq)$. Let $s, t$ be a diametral pair of $T + pc$. Either we have $s, t \in T$ and $pq$ is indifferent for $\{s, t\}$, or we have $s, t \in T$ and $pq$ is useful for $\{s, t\}$, or we have $s \notin T$ or $t \notin T$. We show $\mathrm{diam}(T + pc) \leq \mathrm{diam}(T + pq)$, for the first two cases and we rule out the third case.

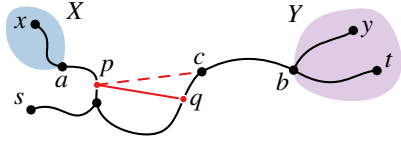 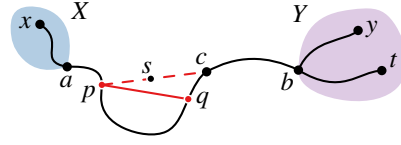

Figure 12: A shortcut $pq$ for $T$ with both endpoints on one side of the absolute center $c$ with a diametral pair $(s, t)$ of $T + pc$ for which $pq$ is useful.

Figure 13: A shortcut $pq$ for $T$ with both endpoints on one side of the absolute center $c$ with an impossible diametral pair $s, t$ of $T + pc$ with $s \notin T$.

1. Suppose $s, t \in T$ and $pq$ is indifferent for $\{s, t\}$, i.e., $d_T(s, t) = d_{T+pq}(s, t)$. Then we have $\text{diam}(T + pc) \leq \text{diam}(T + pq)$, because

   $$\text{diam}(T + pc) = d_{T+pc}(s, t) \leq d_T(s, t) = d_{T+pq}(s, t) \leq \text{diam}(T + pq) \ .$$

2. Suppose $s, t \in T$ and $pq$ is useful for $\{s, t\}$, i.e., $d_{T+pq}(s, t) < d_T(s, t)$.
   Then $pq$ is useful for $(s, t)$ or $(t, s)$. Without loss of generality, let $pq$ be useful for $(s, t)$. Otherwise, we swap $s$ and $t$. Figure 12 illustrates the following arguments. Let $X$ and $Y$ be the primary $\mathcal{B}$-sub-trees with roots $a$ and $b$, respectively, and let $x, y$ be a diametral pair of $T$ with $x \in X$ and $y \in Y$. The shortcut $pq$ is useful for $T$ and, thus, $pq$ is useful for $\{x, y\}$. Hence, $pq$ is useful for $(x, y)$, since $p$ lies on the path from $x$ to $q$ in $T$. The path from $s$ to $p$ cannot contain $q$, since $pq$ is useful for $(s, t)$. This means $s$ lies in the largest sub-tree of $T$ with leaves $x$ and $q$ and the path from $s$ to $c$ in $T$ contains $q$. In the following, we argue $t \in Y$, which implies that $c$ lies on the path from $s$ to $t$ in $T + pc$. As $pq$ is useful for $(s, t)$, the shortcut $pq$ is useful for $(s, q)$, i.e., $d_T(s, p) + |pq| < d_T(s, q)$, since

   $$d_T(s, p) + |pq| + d_T(q, t) = d_{T+pq}(s, t) < d_T(s, t) \leq d_T(s, q) + d_T(q, t) \ .$$

   This means $pc$ is also useful for $(s, c)$, i.e., $d_T(s, p) + |pc| < d_T(s, c)$, since

   $$d_T(s, p) + |pc| \leq d_T(s, p) + |pq| + d_T(q, c) < d_T(s, q) + d_T(q, c) = d_T(s, c) \ .$$

   The point $t$ is a farthest leaf from $c$ in $T$, i.e., $d_T(c, t) = d_T(c, y)$, because

   $$\begin{aligned} d_T(s, p) + |pc| + d_T(c, t) &\leq d_T(s, p) + |pc| + d_T(c, y) \\ &= d_{T+pc}(s, c) + d_T(c, y) \quad (pc \text{ is useful for } (s, c)) \\ &= d_{T+pc}(s, y) \quad (c \text{ lies on the path from } s \text{ to } y) \\ &\leq d_{T+pc}(s, t) \quad (s, t \text{ is diametral in } T + pc) \\ &\leq d_T(s, p) + |pc| + d_T(c, t) \ . \end{aligned}$$

   Since $pq$ was useful for $(s, t)$, the path from $q$ to $t$ cannot contain $p$. On the other hand, $p$ lies on the backbone and blocks the path from $q$ to any farthest leaf from $c$ in $X$. Since $t$ is a farthest leaf from $c$ in $T$, this means that $t \in Y$ and, thus,

   $$\text{diam}(T + pc) = d_{T+pc}(s, t) \qquad (s, t \text{ diametral in } T + pc)$$

$$\leq d_T(s,p) + |pc| + d_T(c,t)$$
$$\leq d_T(s,p) + |pq| + d_T(q,c) + d_T(c,t) \quad \text{(triangle inequality)}$$
$$= d_T(s,p) + |pq| + d_T(q,t) \quad \text{($c$ is on the path from $q$ to $t$)}$$
$$= d_{T+pq}(s,t) \quad \text{($pq$ is useful for $(s,t)$)}$$
$$\leq \operatorname{diam}(T + pq) \ .$$

3. Suppose $s \notin T$ or $t \notin T$. We assume, without loss of generality, that $s \notin T$, i.e., $s \in pc$ with $p \neq s \neq c$. Otherwise, we swap $s$ and $t$.
   Let $X$ and $Y$ be the primary $\mathcal{B}$-sub-trees with roots $a$ and $b$, respectively. As illustrated in Figure 13, the point $s$ lies on the simple cycle $C(p,c)$ in $T + pc$. The largest tree attached to $C(p,c)$ in $T + pc$ is the one containing $Y$, since $c$ is the absolute center of $T$ and both $p$ and $q$ lie on the path from $a$ to $c$. Therefore, the point $t$ lies in $Y$ and $s$ is the farthest point from $t$ on $C(p,c)$. The path from $t$ to $s$ in $T + pc$ enters $C(p,c)$ at $c$. Therefore, $s$ is the farthest point from $c$ on $C(p,c)$. However, this implies that $s$ lies on the path from $p$ to $c$ in $T$, since $|pc| \leq d_T(p,c)$ contradicting $s \notin T$. This means that this case is impossible.

Thus, if $pq$ is an optimal shortcut for $T$ such that $p$ and $q$ lie on the path from $a$ to $c$ in $T$ with $d_T(q,c) \leq d_T(p,c)$ then $pc$ is also an optimal shortcut for $T$. Therefore, there exist an optimal shortcut $pq$ for $T$ with $p,q \in \mathcal{B}$ and $c$ on the path from $p$ to $q$ in $T$. □

## 4. Preparations for the Algorithm

Our search for an optimal shortcut $pq$ for $T$ proceeds as follows. Initially, we place the endpoints of the shortcut, $p$ and $q$, at the absolute center $c$ of $T$. Then, we move $p$ and $q$ along the backbone $\mathcal{B}$ balancing the diametral paths in $T + pq$. Throughout this movement $p$ remains along the path from $a$ to $c$ and $q$ remains on the path from $c$ to $b$, where $a$ and $b$ are the endpoints of $\mathcal{B}$. The diametral pairs in $T + pq$ guide our search: each diametral pair in $T + pq$ rules out some direction in which we could search for a better shortcut. We have found an optimal shortcut when $p$ and $q$ reach a position where the diametral pairs block all directions of movement, except perhaps going back the way we came. We describe our algorithm along the following steps.

1. We simplify the geometric tree $T$ by compressing the $\mathcal{B}$-sub-trees, thereby simplifying the discussion about diametral pairs and paths in $T + pq$.
2. We define algorithm states in terms of the diametral paths and diametral pairs that are present in the augmented tree, and we distinguish four types of movements for the shortcut, called *in-shift*, *out-shift*, *x-shift*, and *y-shift*.
3. We show that each type of diametral pair rules out a better shortcut in some direction, and that some combinations of pair types imply optimality.
4. We describe a continuous and conceptual movement of the shortcut that is guided by the set of types of diametral pairs of $T + pq$. We identify the invariants that are upheld by this movement and that guarantee that we find an optimal shortcut.
5. We specify the speeds at which the endpoints of the shortcut would move in the continuous algorithm. These speeds depend on the set of types of diametral paths in $T + pq$; the changes in this set constitute the events for the discretization.

6. We bound the number of events of the discrete algorithm by $O(n)$. This involves ruling out some transitions between the algorithm states as well as identifying situations where we can safely ignore events without compromising optimality.
7. Finally, we explain how we can process each of the $O(n)$ events in $O(\log n)$ amortized time and, thus, bound the running time of our algorithm by $O(n \log n)$.

In this section, we discuss Steps 1, 2, and 3, i.e., the preparations for the algorithm. In Section 5, we describe Step 4, i.e., the continuous algorithm and its correctness. In Section 6, we discuss Steps 5 through 7, i.e., the discretization of the algorithm.

*4.1. Simplifying the Tree*

Let $T$ be a geometric tree whose backbone $B$ consists of more than its absolute center $c$. Let $a$ and $b$ be the endpoints of $\mathcal{B}$, and let $X$ and $Y$ be the primary $\mathcal{B}$-sub-trees of $T$ with roots $a$ and $b$, respectively, and let $S_1, S_2, \ldots, S_k$ be the secondary $\mathcal{B}$-sub-trees of $T$ that are attached to $\mathcal{B}$ at their roots $r_1, r_2, \ldots, r_k$, respectively. We assume, without loss of generality, that the roots $r_1, r_2, \ldots, r_k$ appear in this order from $a$ to $b$ along $\mathcal{B}$. Let $x$ be a farthest leaf from $a$ in $X$, let $y$ be a farthest leaf from $b$ in $Y$ and, for every $i = 1, 2, \ldots, k$, let $s_i$ be a farthest leaf from $r_i$ in $S_i$, as illustrated in Figure 14a.

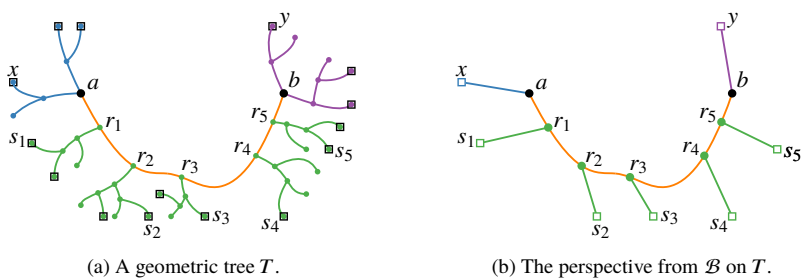

(a) A geometric tree $T$.  (b) The perspective from $\mathcal{B}$ on $T$.

Figure 14: A geometric tree a together with the perspective from its backbone b. We represent each $\mathcal{B}$-sub-tree $S$ with an edge whose length is the length of a path from the root of $S$ to a farthest leaf (squares) of $S$.

We simplify the discussion about diametral pairs in $T + pq$. First, there is no need to distinguish diametral pairs with partners in the same $\mathcal{B}$-sub-trees: for any $i, j = 1, 2, \ldots, k$ with $i \neq j$, the pair $s_i, s_j$ is diametral in $T + pq$ if and only if every farthest leaf from $r_i$ in $S_i$ forms a diametral pair with every farthest leaf from $r_j$ in $S_j$. Second, there is no need to consider diametral pairs with both endpoints in the same $\mathcal{B}$-sub-tree, as we argue in Lemma 8 below. Therefore, we simplify $T$ by replacing each $\mathcal{B}$-sub-tree $S_i$ with an edge from $r_i$ to a vertex representing $s_i$ of length $d_T(r_i, s_i)$. Likewise, we replace $X$ and $Y$ with edges of appropriate length, as illustrated in Figure 14. We refer to the resulting caterpillar network as the *perspective* from $\mathcal{B}$ on $T$.

**Lemma 8** *Let $T$ be a geometric tree with backbone $\mathcal{B}$, let $p, q \in \mathcal{B}$, and let $S$ be a $\mathcal{B}$-sub-tree of $T$. If there exist $u, v \in S$ such that $u, v$ is diametral in $T + pq$, then $pq$ is an optimal shortcut for $T$.*

PROOF. Let $T$ be a geometric tree with backbone $\mathcal{B}$, let $p, q \in \mathcal{B}$, and let $S$ be a $\mathcal{B}$-sub-tree of $T$. Suppose there exist $u, v \in S$ such that $u, v$ is a diametral pair of $T + pq$.

Since $S$ is a tree that is attached to the remainder of $T + pq$, we have $\operatorname{diam}(S) \leq \operatorname{diam}(T + pq)$. Every path from $u$ to $v$ via $pq$ contains $r$ twice, hence the shortest path from $u$ to $v$ in $T + pq$ remains in $S$. This implies $\operatorname{diam}(T + pq) = \operatorname{diam}(S) = d_S(u, v)$, since $\operatorname{diam}(T + pq) = d_{T+pq}(u, v) = d_S(u, v) \leq \operatorname{diam}(S) \leq \operatorname{diam}(T + pq)$. By Lemma 6, there is an optimal shortcut $p^*q^*$ for $T$ with $p^*, q^* \in \mathcal{B}$. By repeating the above, we obtain $d_S(u, v) = d_{T+p^*q^*}(u, v)$ and, thus, $\operatorname{diam}(T + pq) = \operatorname{diam}(T + p^*q^*)$, since

$$\operatorname{diam}(T + p^*q^*) \leq \operatorname{diam}(T + pq) = \operatorname{diam}(S)$$
$$= d_S(u, v) = d_{T+p^*q^*}(u, v) \leq \operatorname{diam}(T + p^*q^*) \ .$$

Therefore, $pq$ is an optimal shortcut for $T$ and $\operatorname{diam}(T + pq) = \operatorname{diam}(S)$. □

**Corollary 9** *Let $T$ be a geometric tree, and let $\delta$ be the largest diameter of any $\mathcal{B}$-sub-tree of $T$. For every shortcut $pq$ for $T$, we have $\delta \leq \operatorname{diam}(T + pq)$.* □

We compute the largest diameter $\delta$ of any $\mathcal{B}$-sub-tree of $T$ as part of our preprocessing and we halt our search for an optimal shortcut if the current diameter reaches $\delta$. This is not strictly necessary: if we ignore diametral pairs in the same $\mathcal{B}$-sub-tree, we still obtain an optimal shortcut even though we might not know the true optimal diameter. We may safely exclude diametral pairs in the same $\mathcal{B}$-sub-tree from consideration.

*4.2. States and Operations*

We group the diametral pairs and paths of the augmented tree $T + pq$ into types. The types of diametral pairs and paths in $T + pq$ define the states of our algorithm. We specify four types of movements for the shortcut as the operations for the algorithm.

*4.2.1. Pair States*

We define the following symbols make the discussion about the candidates for diametral pairs of the augmented tree $T + pq$ more accessible.

- *x*: A point in the primary $\mathcal{B}$-sub-tree $X$.
- *y*: A point in the primary $\mathcal{B}$-sub-tree $Y$.
- ■: A point on $T \setminus (X \cup Y)$.
- ▲: A point in a secondary $\mathcal{B}$-sub-tree $S_i$ for some $i = 1, 2, \ldots, k$.
- ●: A point on the simple cycle $C(p, q)$ in $T + pq$ that lies on $T$.
- ○: A point on the simple cycle $C(p, q)$ in $T + pq$ that lies on $pq$.

Every point on $T + pq$ lies (*x*) in the primary $\mathcal{B}$-sub-tree $X$, or (*y*) in the primary $\mathcal{B}$-sub-tree $Y$, or (■) somewhere along $T \setminus (X \cup Y)$, or (○) on the shortcut $pq$.

If $u, v$ is a diametral pair of $T + pq$, then $u$ is $(x, y)$ a leaf of a primary $\mathcal{B}$-sub-tree, or (▲) a leaf of a secondary $\mathcal{B}$-sub-tree, or (●, ○) $u$ is a point along the simple cycle $C(p, q)$ in $T + pq$. If $u$ lies on the cycle $C(p, q)$, then $u$ lies (●) on $T$ or (○) on the shortcut $pq$. This distinction is complete: the only remaining location for $u$ would be along a part of the backbone that does not belong to the cycle $C(p, q)$, i.e., on the path from $a$ to $p$ or the path from $q$ to $b$. In both cases, we could extend the supposedly diametral path from $v$ to $u$ in $T + pq$ to a strictly longer path from $v$ to a leaf of $X$ or a leaf of $Y$.

This leads to the following types of diametral pairs in $T + pq$. The choice for this particular distinction will become clear when we discuss the details of the algorithm.

**$x$-$y$:** Diametral pairs $x, y$ of $T + pq$ with $x \in X$ and $y \in Y$.

**$x$-■:** Diametral pairs $x, v$ of $T + pq$ with $x \in X$ and $v \in T \setminus (X \cup Y)$.

Diametral pairs of this type manifest as one of the following two sub-types.

**$x$-▲:** Diametral pairs $x, s_j$ of $T + pq$ with $x \in X$ and $s_j \in S_j$ for $j = 1, 2, \ldots, k$.

**$x$-●:** Diametral pairs $x, \bar{x}$ of $T + pq$, where $x \in X$ and where $\bar{x}$ is the farthest point from $x$ on $C(p, q)$. Since $|pq| \le d_T(p, q)$, we always have $\bar{x} \in T$.

**■-$y$:** Diametral pairs $u, y$ of $T + pq$ with $u \in T \setminus (X \cup Y)$ and $y \in Y$.

Diametral pairs of this type manifest as one of the following two sub-types.

**▲-$y$:** Diametral pairs $s_i, y$ of $T + pq$ with $s_i \in S_i$ and $y \in Y$ for some $i = 1, 2, \ldots, k$.

**●-$y$:** Diametral pairs $y, \bar{y}$ of $T + pq$, where $y \in Y$ and where $\bar{y}$ is the farthest point from $y$ on $C(p, q)$. Since $|pq| \le d_T(p, q)$, we always have $\bar{y} \in T$.

**■-■:** Diametral pairs $u, v$ of $T + pq$ with $u, v \in T \setminus (X \cup Y)$.

Diametral pairs of this type manifest as one of the following two sub-types.

**▲-▲:** Diametral pairs $s_i, s_j$ of $T + pq$ with $s_i \in S_i$ and $s_j \in S_j$ for $i, j = 1, 2, \ldots, k$.

**▲-●:** Diametral pairs $s_i, \bar{s}_i$ of $T + pq$ with $s_i, \bar{s}_i \in T$ where $s_i \in S_i$ for some $i = 1, 2, \ldots, k$ and where $\bar{s}_i \in T$ is the farthest point from $s_i$ on $C(p, q)$.

**■-○:** Diametral pairs $u, v$ of $T + pq$ with $v \notin T$, i.e., $v \in pq$ with $p \ne v \ne q$.

Since $|pq| \le d_T(p, q)$, we have $u \in T$. These diametral pairs only manifest as sub-type ▲-○, i.e., in the form $s_i, \bar{s}_i$ with $s_i \in S_i$, for $i = 1, 2, \ldots, k$, and where $\bar{s}_i$ is the farthest point from $s_i$ on $C(p, q)$ that lies in the interior of the shortcut $pq$.

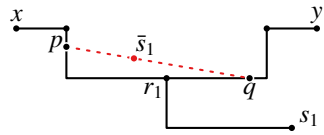

Figure 15: An optimal shortcut $pq$ for a geometric tree $T$. The diametral pairs in the augmented tree $T + pq$ are $(x, y)$, $(x, s_1)$, $(s_1, y)$, and $(\bar{s}_1, s_1)$, i.e., $T + pq$ is in pair state $\{x\text{-}y, x\text{-}■, ■\text{-}y, ■\text{-}○\}$ and in pair sub-state $\{x\text{-}y, x\text{-}▲, ▲\text{-}y, ▲\text{-}○\}$. Since $pq$ is useful for $x, y$, but useless for $(x, s_1)$ and $(s_1, y)$, the path state of $T + pq$ is $\{x\text{-}pq\text{-}y, x\text{-}T\text{-}▲, ▲\text{-}T\text{-}y, ▲\text{-}p\text{-}○, ▲\text{-}q\text{-}○\}$.

The order of the types of endpoints in the above notation is entirely arbitrary, e.g., $x$-■ and ■-$x$ describe the same type of pair. There is no need to consider diametral pairs of sub-type ●-● or ●-○: the distance from $x$ or from $y$ to their farthest points on $C(p, q)$ is always larger than the distance between any two points on $C(p, q)$—unless $T$ is a path with endpoints $p$ and $q$, i.e., $T + pq = C(p, q)$. Since $|pq| \le d_T(p, q)$, the farthest points $\bar{x}$ and $\bar{y}$ from $x$ and from $y$ on $C(p, q)$ lie on $T$. If $x, v$ is a diametral pair of type $x$-■ in $T + pq$, then $v \in T$, and if $u, y$ is of type ■-$y$ in $T + pq$, then $u \in T$. Therefore, there do not exist any diametral pairs of sub-type $x$-○ or ○-$y$.

The *pair state* is the set of types of diametral pairs in $T + pq$ and the *pair sub-state* is the set of sub-types of diametral pairs that are present in $T + pq$. For instance, if $T + pq$ has the diametral pairs $x, y$; $x, s_3$; $x, s_5$; and $x, \bar{x}$ then $T + pq$ is in pair state $\{x\text{-}y, x\text{-}\blacksquare\}$ and in the pair sub-state $\{x\text{-}y, x\text{-}\blacktriangle, x\text{-}\bullet\}$. Figure 15 illustrates the pair state and pair sub-state of an augmented tree together with its path state that we introduce next.

*4.2.2. Path States*

Let $u, v$ be a diametral pair in $T + pq$. If $u, v \in T$ then every diametral path in $T + pq$ that connects $u$ and $v$ either contains the shortcut ($*$-$pq$-$*$) or not ($*$-$T$-$*$). If $u \in T$ and $v \notin T$, then every diametral path in $T + pq$ that connects $u$ and $v$ contains either $p$ ($*$-$p$-$*$) or $q$ ($*$-$q$-$*$). There are no diametral pairs with $u, v \notin T$. This leads to the following.

- ($*$-$\boldsymbol{pq}$-$*$): A diametral path that *does* contain $pq$ and connects $u \in T$ with $v \in T$.
- ($*$-$\boldsymbol{T}$-$*$): A diametral path that *does not* contain $pq$ and connects $u \in T$ with $v \in T$.
- ($*$-$\boldsymbol{p}$-$*$): A diametral path that contains $p$ and connects $u \in T$ with $v \notin T$.
- ($*$-$\boldsymbol{q}$-$*$): A diametral path that contains $q$ and connects $u \in T$ with $v \notin T$.

Any type of diametral partner (e.g., $x, y, \blacksquare, \blacktriangle, \bullet, \circ$) may appear in place of $*$. For instance, we denote a diametral path from $x$ to $\bar{x}$ via the shortcut by $x$-$pq$-$\bullet$. The *path state* is the set of types of diametral paths that are present in $T + pq$. If $T + pq$ has the diametral pairs $x, y$; $x, s_3$; $x, s_5$; and $x, \bar{x}$ such that $pq$ is useful for $(x, y)$ and $(x, s_3)$ but useless for $(x, s_5)$ then $T + pq$ is in path state $\{x\text{-}pq\text{-}y, x\text{-}pq\text{-}\blacktriangle, x\text{-}T\text{-}\blacktriangle, x\text{-}pq\text{-}\bullet, x\text{-}T\text{-}\bullet\}$.

*4.2.3. Operations*

We distinguish the four types of movements for the shortcut depicted in Figure 16. Suppose we move a shortcut $pq$ with $p, q \in \mathcal{B}$ such that $d_T(a, p) \leq d_T(a, q)$ to a position $p'q'$ with $p', q' \in \mathcal{B}$ such that $d_T(a, p') \leq d_T(a, q')$. The movement from $pq$ to $p'q'$ is

- an *out-shift* if $p$ approaches $a$ and $q$ approaches $b$, i.e., $d_T(a, p') \leq d_T(a, p)$ and $d_T(b, q') \leq d_T(b, q)$ and $d_T(a, p') \leq d_T(a, p) \leq d_T(a, q) \leq d_T(a, q')$,

- an *in-shift* if $p$ recedes from $a$ and $q$ recedes from $b$, i.e., $d_T(a, p) \leq d_T(a, p')$ and $d_T(b, q) \leq d_T(b, q')$ and $d_T(a, p) \leq d_T(a, p') \leq d_T(a, q') \leq d_T(a, q)$,

- an *x-shift* if $p$ approaches $a$ and $q$ recedes from $b$, i.e., $d_T(a, p') \leq d_T(a, p)$ and $d_T(b, q) \leq d_T(b, q')$ and $d_T(a, p') \leq d_T(a, p) \leq d_T(a, q') \leq d_T(a, q)$, or

- a *y-shift* if $p$ recedes from $a$ and $q$ approaches $b$, i.e., $d_T(a, p) \leq d_T(a, p')$ and $d_T(b, q') \leq d_T(b, q)$ and $d_T(a, p) \leq d_T(a, p') \leq d_T(a, q) \leq d_T(a, q')$.

These types of movements intentionally overlap when one of the endpoints remains stationary, e.g., when $p = p'$ every y-shift is also an out-shift, as illustrated in Figure 16e. An out-shift is short for an outward shift, an in-shift is short for an inward shift, an x-shift is short for a shift towards $x$, and a y-shift is short for a shift towards $y$.

The operations, as defined above, allow us to move $p$ or $q$ through the absolute center $c$. However, the algorithm automatically maintains that $p$ stays on the path from $a$ to $c$ and $q$ stays on the path from $c$ to $b$—without taking any special care to ensure this.

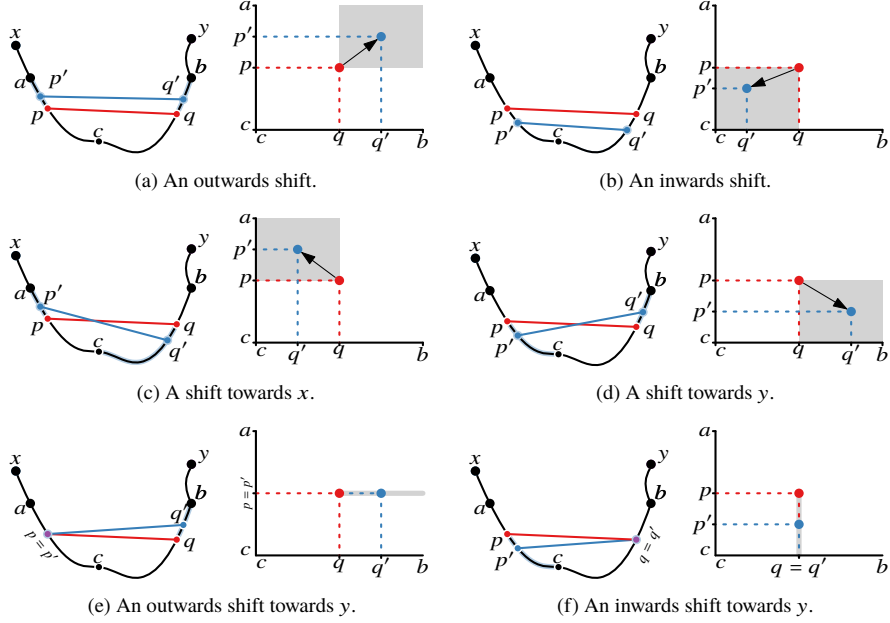

Figure 16: The four ways to move a shortcut $pq$ to a new position $p'q'$. They are (a) the in-shift, (b) the out-shift, (c) the $x$-shift, and (d) the $y$-shift. These cases intentionally overlap when one endpoint of the shortcut remains stationary, i.e., when $p = p'$, as in (e), or when $q = q'$, as in (f). We sketch each operation on the left and plot of the positions of the shortcuts along the path from $c$ to $a$ and the path from $c$ to $b$ on the right. In each plot, the shaded region indicates all the shortcuts that can be reached with the same movement.

*4.3. Blocking*

Each type of diametral pair certifies that some type of movement cannot lead to a better shortcut, i.e., cannot reduce the continuous diameter. For instance, if an augmented tree $T + pq$ has a diametral pair of type $x$-$y$, then the distance between $x$ and $y$ will increase or remain the same when we in-shift $pq$. In this sense, $x$-$y$ *blocks* any in-shift, $x$-■ *blocks* any $y$-shift, ■-$y$ *blocks* any $x$-shift, and ■-■ and ■-○ *block* any out-shift.

**Lemma 10 (Blocking)** *Let $pq$ be a shortcut for a geometric tree $T$ with $p, q \in \mathcal{B}$.*

(1) *If $T + pq$ has a diametral pair of type $x$-$y$, then $\mathrm{diam}(T + pq) \leq \mathrm{diam}(T + p'q')$ for every shortcut $p'q'$ such that the movement from $pq$ to $p'q'$ is an inward shift.*

(2) *If $T + pq$ has a diametral pair of type $x$-■, then $\mathrm{diam}(T + pq) \leq \mathrm{diam}(T + p'q')$ for every shortcut $p'q'$ such that the movement from $pq$ to $p'q'$ is a shift towards $y$.*

(3) *If $T + pq$ has a diametral pair of type ■-$y$, then $\mathrm{diam}(T + pq) \leq \mathrm{diam}(T + p'q')$ for every shortcut $p'q'$ such that the movement from $pq$ to $p'q'$ is a shift towards $x$.*

(4) *If $T + pq$ has a diametral pair of type ■-■ or ■-○, then $\mathrm{diam}(T + pq) \leq \mathrm{diam}(T + p'q')$ for every shortcut $p'q'$ such that the movement from $pq$ to $p'q'$ is an outward shift.*

PROOF. Let $T$ be a geometric tree with backbone $\mathcal{B}$, and let $p, q, p', q' \in \mathcal{B}$.

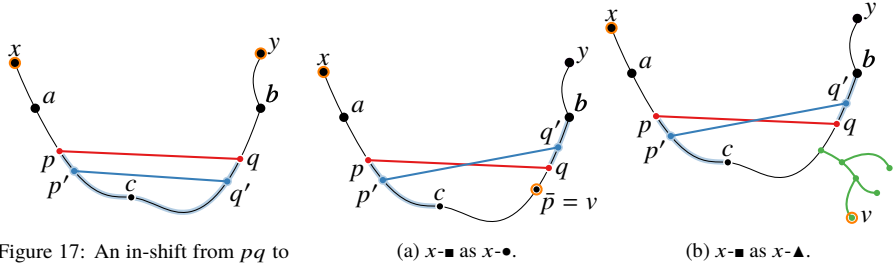

Figure 17: An in-shift from $pq$ to $p'q'$ that increases the distance between a diametral pair $x, y$ of $T+pq$.

(a) $x$-■ as $x$-●.

(b) $x$-■ as $x$-▲.

Figure 18: A $y$-shift from $pq$ to $p'q'$ that increases the distance between a diametral pair $x, v$ of type $x$-■ manifesting as (a) $x$-● and (b) $x$-▲.

If there exists a diametral pair $u, v$ of $T + pq$ such that $u, v \in T$, and $p'q'$ is indifferent for $\{u, v\}$, i.e., $d_T(u, v) = d_{T+p'q'}(u, v)$, then $\mathrm{diam}(T + pq) \leq \mathrm{diam}(T + p'q')$, since

$$\mathrm{diam}(T + pq) = d_{T+pq}(u,v) \leq d_T(u,v) = d_{T+p'q'}(u,v) \leq \mathrm{diam}(T + p'q') \ .$$

Thus, we may assume $p'q'$ is useful for any diametral pair $(u, v)$ of $T + pq$ with $u, v \in T$.

We have $\mathrm{diam}(T + pq) \leq \mathrm{diam}(T + p'q')$ if there exists a diametral pair $u, v$ of $T + pq$ with $u, v \in T$ such that $p'q'$ is useful for $(u, v)$, the path from $u$ to $p'$ in $T$ contains $p$, i.e., $d_T(u, p') = d_T(u, p) + d_T(p, p')$, and the path from $v$ to $q$ contains $q'$, i.e., $d_T(q', v) = d_T(q', q) + d_T(q, v)$, since

$$\begin{aligned}
\mathrm{diam}(T + pq) &= d_{T+pq}(u, v) &&(u, v \text{ is diametral in } T + pq) \\
&\leq d_T(u, p) + |pq| + d_T(q, v) \\
&\leq d_T(u, p) + d_T(p, p') + |p'q'| + d_T(q', q) + d_T(q, v) \\
&= d_T(u, p') + |p'q'| + d_T(q', v) &&(p' \text{ lies on the path from } u \text{ to } p) \\
&= d_{T+p'q'}(u, v) &&(p'q' \text{ is useful for } (u, v)) \\
&\leq \mathrm{diam}(T + p'q') \ .
\end{aligned}$$

We use this observation to prove that each diametral pair blocks the stated operation.

(1) Suppose $T + pq$ has a diametral pair of type $x$-$y$. Let $p'q'$ be a shortcut for $T$ such that the movement from $pq$ to $p'q'$ is an inward shift, as illustrated in Figure 17. Then $\mathrm{diam}(T + pq) \leq \mathrm{diam}(T + p'q')$, since $x, y$ is a diametral pair of $T + pq$ with $x, y \in T$ such that $p$ lies on the path from $x$ to $p'$ and $q$ lies on the path from $q'$ to $y$.

(2) Suppose $T + pq$ has a diametral pair of type $x$-■. Let $p'q'$ be a shortcut for $T$ such that the movement from $pq$ to $p'q'$ is a shift towards $y$, as illustrated in Figure 18. Let $x, v$ be a diametral pair of type $x$-■ in $T + pq$. Then $v \in T$, since $|pq| \leq d_T(p, q)$. The path from $x$ to $p'$ contains $p$, since the movement from $pq$ to $p'q'$ is a shift towards $y$. The path from $q'$ to $v$ in $T$ contains $q$, since otherwise $y$ would be farther away from $x$ than $v$ in $T + pq$. Therefore, we have $\mathrm{diam}(T + pq) \leq \mathrm{diam}(T = p'q')$.

(3) Suppose $T + pq$ has a diametral pair of type ■-$y$. This is symmetric to Case (2).

(4) Suppose $T + pq$ has a diametral pair of type ■-■ or ■-○. Let $p'q'$ be a shortcut for $T$ such that the movement from $pq$ to $p'q'$ is an out-shift, as illustrated in Figure 19. Every diametral pair of type ■-■ or ■-○ manifests as subtype ▲-▲, ▲-● or ▲-○.

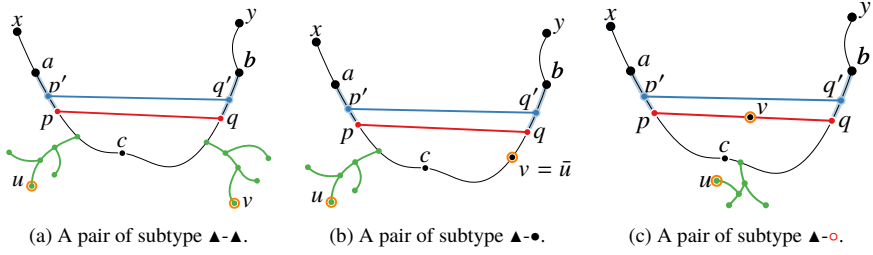

(a) A pair of subtype ▲-▲.     (b) A pair of subtype ▲-●.     (c) A pair of subtype ▲-○.

Figure 19: An illustration of an outward shift from $pq$ to $p'q'$ with a diametral pair $u, v$ in $T + pq$ manifesting as (a) subtype ▲-▲, (b) subtype ▲-●, and (c) subtype ▲-○.

(a) Suppose there is some diametral pair $u, v$ of $T + pq$ of subtype ▲-▲. Then $u$ and $v$ are leaves of two secondary $\mathcal{B}$-sub-trees that are attached to $\mathcal{B}$ along the path from $p$ to $q$ in $T$. Otherwise, $x$ or $y$ would be strictly farther from $v$ than $u$. Then we have $\mathrm{diam}(T + pq) \leq \mathrm{diam}(T + p'q')$, since $u, v$ is a diametral pair of $T + pq$ with $u, v \in T$ where $p$ lies on the path from $u$ to $p'$ and $q$ lies on the path from $q'$ to $v$.

(b) Suppose there is some diametral pair $u, v$ in $T + pq$ of subtype ▲-● or ▲-○. Then one of $u$ or $v$ is a leaf of a secondary $\mathcal{B}$-sub-tree and the other is the farthest point on $C(p, q)$ from said leaf. Without loss of generality, suppose $u$ lies in a secondary $\mathcal{B}$-sub-tree $S$. The root $r$ of $S$ must lie along the path from $p$ to $q$ in $T$. Otherwise, $x$ or $y$ would be strictly farther from $u$ than $v$. Thus,

$$d_{T+pq}(u, v) = d_T(u, r) + d_{T+pq}(r, v) = d_T(u, r) + \frac{|pq| + d_T(p, q)}{2} \ .$$

Let $v'$ be the farthest point from $u$ on the simple cycle $C(p'q')$ in $T + p'q'$. The vertex $r$ lies on the path from $p'$ to $q'$ in $T$, since the movement from $pq$ to $p'q'$ is an outward shift, and since $r$ lies on the path from $p$ to $q$ in $T$. Thus,

$$d_{T+p'q'}(u, v') = d_T(u, r) + d_{T+p'q'}(r, v') = d_T(u, r) + \frac{|p'q'| + d_T(p', q')}{2} \ .$$

The simple cycle in the augmented tree is growing as $pq$ shifts outwards to $p'q'$, i.e., $|pq| + d_T(p, q) \leq |p'q'| + d_T(p'q')$. This implies $d_{T+pq}(u, v) \leq d_{T+p'q'}(u, v')$ and, therefore, $\mathrm{diam}(T + pq) = d_{T+pq}(u, v) \leq d_{T+p'q'}(u, v') \leq \mathrm{diam}(T + p'q')$.

Therefore, each type of diametral pair blocks one type of movement, as claimed. □

*4.4. Sudden Optimality*

As a consequence of Lemma 10, a shortcut $pq$ is optimal for a geometric tree $T$ when each of the four types of movements is blocked by some diametral pair in $T + pq$.

**Corollary 11** *If an augmented tree $T + pq$ has diametral pairs of type x-y, x-■, and ■-y, as well as a diametral pair of type ■-■ or ■-○, then $pq$ is an optimal shortcut for $T$.* □

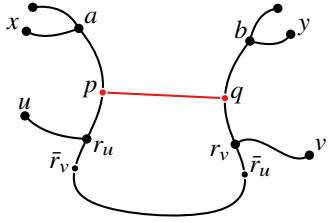 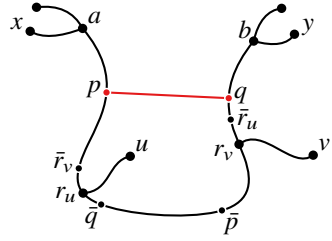

Figure 20: An augmented tree $T + pq$ with diametral paths of type $x$-$pq$-$y$ and ■-$pq$-■. The diametral path of type ■-$pq$-■ connects $u$ with $v$ via $pq$. The points $\bar{r}_u$ and $\bar{r}_v$ mark the farthest points from $r_u$ and $r_v$, respectively on the simple cycle in $T + pq$. Their position indicates that $pq$ is useful for $(u, v)$.

Figure 21: An augmented tree $T + pq$ with diametral paths of type $x$-$pq$-$y$ and ■-$T$-■. The diametral path of type ■-$T$-■ connects $u$ with $v$ via the tree only. The points $\bar{r}_u$ and $\bar{r}_v$ mark the farthest points from $r_u$ and $r_v$, respectively on the simple cycle in $T + pq$. Their position indicates that $pq$ is useless for $(u, v)$.

We argue that we may ignore diametral pairs of type ■-■ when a diametral pair of $x$-$y$ is present, provided that we keep track of diametral pairs of type ■-$y$ and $x$-■. This is important for the running time, as there may be $\Omega(n^2)$ candidates for pairs of type ■-■.

**Theorem 12** *If an augmented tree $T + pq$ has diametral pairs of types $x$-$y$ and ■-■, then $T + pq$ has diametral pairs of type $x$-■ and ■-$y$ and, thus, $pq$ is optimal for $T$.*

PROOF. Suppose an augmented tree $T + pq$ has diametral pairs of types $x$-$y$ and ■-■. Then $pq$ is useful for $(x, y)$, since $T + pq$ has diametral pairs besides those of type $x$-$y$. Therefore, the diametral pairs of type $x$-$y$ in $T + pq$ are connected by diametral paths of type $x$-$pq$-$y$. We distinguish whether the diametral pairs of type ■-■ in $T + pq$ are connected by diametral paths of type ■-$pq$-■ or ■-$T$-■. In each case, we argue that there are diametral pairs of types $x$-■ and of type ■-$y$, which implies that $pq$ is optimal for $T$.

1. Suppose $T + pq$ has diametral paths of type $x$-$pq$-$y$ and ■-$pq$-■, as in Figure 20. Then $T + pq$ has a diametral pair $u, v \in T \setminus (X \cup Y)$ such that a shortest path from $u$ to $v$ contains $pq$. We show that the pairs $u, y$ and $x, v$ are diametral in $T + pq$. Since $pq$ is useful for $(u, v)$, the shortcut $pq$ is also useful for $(u, q)$ an, thus, $(u, y)$. By comparing the paths $x$-$pq$-$y$ and $u$-$pq$-$y$, we obtain $d_T(u, p) \leq d_T(x, p)$, since

$$d_T(u, p) + |pq| + d_T(q, y) = d_{T+pq}(u, y) \qquad (pq \text{ is useful for } (u, y))$$
$$\leq d_{T+pq}(x, y) \qquad (x, y \text{ is diametral in } T + pq)$$
$$= d_T(x, p) + |pq| + d_T(q, y) \ .$$
$$\qquad (pq \text{ is useful for } (x, y))$$

Analoguously, we obtain $d_T(q, v) \leq d_T(q, y)$ by comparing the paths $x$-$pq$-$y$ and $x$-$pq$-$v$. Comparing the diametral paths $x$-$pq$-$y$ and $u$-$pq$-$v$ yields $d_T(x, p) + d_T(q, y) = d_T(u, p) + d_T(q, v)$. This equation cannot be satisfied when $d_T(u, p) < d_T(x, p)$ or $d_T(q, v) < d_T(q, y)$, since $d_T(u, p) \leq d_T(x, p)$ and $d_T(q, v) \leq d_T(q, y)$. Therefore, we have $d_T(u, p) = d_T(x, p)$ and $d_T(q, v) = d_T(q, y)$. This implies that $u, y$ and $x, v$ are diametral pairs in $T + pq$, since

$$d_{T+pq}(u, y) = d_T(u, p) + |pq| + d_T(q, y)$$

$$= d_T(x,p) + |pq| + d_T(q,y) = \text{diam}(T+pq) \ ,$$
$$\text{and} \ \ d_{T+pq}(x,v) = d_T(x,p) + |pq| + d_T(q,v)$$
$$= d_T(x,p) + |pq| + d_T(q,y) = \text{diam}(T+pq) \ .$$

Hence, $T + pq$ is in the pair state $\{x\text{-}y, \blacksquare\text{-}\blacksquare, x\text{-}\blacksquare, \blacksquare\text{-}y\}$ and, thus, $pq$ is optimal for $T$.

2. Suppose $T + pq$ has diametral paths of type $x\text{-}pq\text{-}y$ and $\blacksquare\text{-}T\text{-}\blacksquare$, as in Figure 21. Then there exists a diametral pair $u, v$ of $T + pq$ with $u, v \in T$ such that there is a shortest path from $u$ to $v$ in $T + pq$ that does not contain $pq$. If $u$ lies in a secondary $\mathcal{B}$-sub-tree $S_u$, then let $r_u$ be the root of $S_u$. Otherwise, let $r_u = u$. Likewise, let $r_v$ be the root of the $\mathcal{B}$-sub-tree containing $v$ or let $r_v = v$ when $v \in \mathcal{B}$. Without loss of generality, $r_u$ lies on the path in $T$ from $a$ to $r_v$. Otherwise, we swap $u$ and $v$. We show that $u, y$ and $x, v$ are diametral in $T + pq$. Since $u\text{-}T\text{-}v$ is diametral, $pq$ cannot be useful for $(u, v)$ and, thus, $pq$ cannot be useful for $(r_u, r_v)$, i.e., $d_T(r_u, r_v) \leq d_T(r_u, p) + |pq| + d_T(q, r_v)$. On the other hand, $pq$ must be useful for $(x, v)$, i.e., $d_{T+pq}(x, v) < d_T(x, v)$, since

$$\begin{aligned}
d_{T+pq}(x, v) &\leq \text{diam}(T + pq) \\
&= d_{T+pq}(u, v) &&(u, v \text{ is diametral in } T + pq) \\
&= d_T(u, v) &&(pq \text{ is not useful for } (u, v)) \\
&= d_T(u, r_u) + d_T(r_u, v) &&(r_u \text{ lies on the path from } u \text{ to } v) \\
&< d_T(x, r_u) + d_T(r_u, v) &&(u \in T \setminus X) \\
&= d_T(x, v) \ . &&(r_u \text{ lies on the path from } x \text{ to } v)
\end{aligned}$$

Likewise, $pq$ must be useful for $(u, y)$. We have $d_T(q, v) \leq d_T(q, y)$, as

$$\begin{aligned}
d_T(x, p) + |pq| + d_T(q, v) &= d_{T+pq}(x, v) &&(pq \text{ is useful for } (x, v)) \\
&\leq d_{T+pq}(x, y) &&(x, y \text{ is diametral in } T + pq) \\
&= d_T(x, p) + |pq| + d_T(q, y) &&(pq \text{ is useful for } (x, y))
\end{aligned}$$

The pair $u, y$ is diametral in $T + pq$, i.e., $d_{T+pq}(u, y) = \text{diam}(T + pq)$, since

$$\begin{aligned}
\text{diam}(T + pq) &= d_{T+pq}(u, v) &&(u, v \text{ is diametral in } T + pq) \\
&= d_T(u, v) &&(pq \text{ is not useful for } (u, v)) \\
&= d_T(u, r_u) + d_T(r_u, r_v) + d_T(r_v, v) \\
&\leq d_T(u, r_u) + d_T(r_u, p) + |pq| + d_T(q, r_v) + d_T(r_v, v) \\
& &&(pq \text{ not useful for } (r_u, r_v)) \\
&= d_T(u, p) + |pq| + d_T(q, v) \\
&\leq d_T(u, p) + |pq| + d_T(q, y) &&(d_T(q, v) \leq d_T(q, y)) \\
&= d_{T+pq}(u, y) \leq \text{diam}(T + pq) \ . &&(pq \text{ is useful for } (u, y))
\end{aligned}$$

Likewise, $x, v$ is diametral in $T + pq$ and $T + pq$ is in pair state $\{x\text{-}y, x\text{-}\blacksquare, \blacksquare\text{-}y, \blacksquare\text{-}\blacksquare\}$.

Hence, if $T + pq$ has diametral pairs of type $x\text{-}y$ and $\blacksquare\text{-}\blacksquare$, then $pq$ is optimal for $T$. □

## 5. Continuous Algorithm

Inspired by the plane-sweep paradigm, we—conceptually—move the shortcut continuously while changing its speed and direction at certain events where the pair state or path state changes. To implement this approach, we discretize this movement such that the shortcut jumps from one event to the next.

### 5.1. The Algorithm from the Pair State Perspective

Figure 22 describes the continuous algorithm in terms of the pair states and operations. Initially, we place the shortcut with both endpoints on the absolute center $c$ of the geometric tree $T$. This ensures that we start in pair state $\{x\text{-}y\}$. The algorithm consists of at most three phases: an outwards shift, possibly followed by a shift towards $x$ or a shift towards $y$, possibly followed by another outwards shift. Some pair states are marked as final states with a double border. If we reach a final state, we terminate our search and report the best shortcut that we have found. For the other states, we specify the direction in which we move the shortcut.

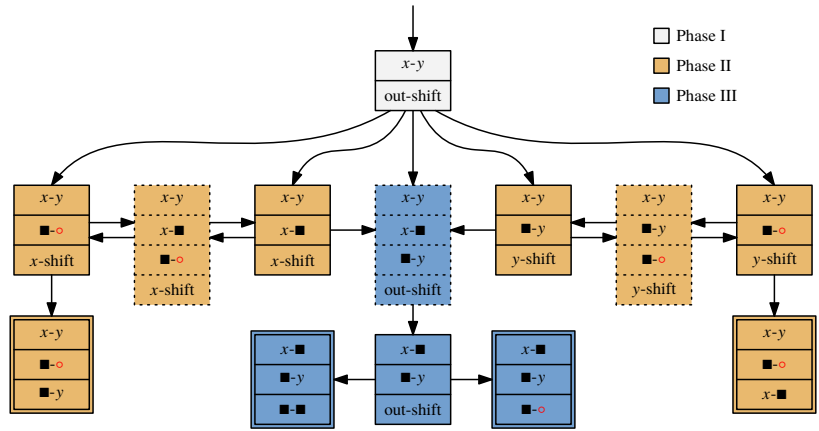

Figure 22: The pair states encountered during our search for an optimal shortcut for a tree. There are three types of states: First, regular states (single boundary) indicate the pair state and the operation applied (out-shift, $x$-shift, or $y$-shift). Second, transition states (dotted boundary) are visited only momentarily while transit from one regular state to another regular state. Third, final states (double boundary) where we terminate our search and report the best shortcut encountered. Under certain conditions, the search may also terminate early in non-final states. We always start in state $\{x\text{-}y\}$ with an outward shift. When we reach the pair state $\{x\text{-}y, \blacksquare\text{-}\circ\}$ then we perform both a shift towards $x$ and separately a shift towards $y$.

For the sake of simplicity, we omit some pair states and transitions from Figure 22. First, we omit all pair states containing $x\text{-}y$ and $\blacksquare\text{-}\blacksquare$, due to Theorem 12. Second, we omit transitions that are implied by transitivity, e.g., we model a transition from $\{x\text{-}y\}$ to $\{x\text{-}y, x\text{-}\blacksquare, \blacksquare\text{-}y\}$ by transitioning from $\{x\text{-}y\}$ to $\{x\text{-}y, x\text{-}\blacksquare\}$ and then from $\{x\text{-}y, x\text{-}\blacksquare\}$ to $\{x\text{-}y, x\text{-}\blacksquare, \blacksquare\text{-}y\}$. Third, we omit pair states that are supersets of any final states.

*Phase I: Shifting Outwards.* In Phase I, we continuously shorten all diametral paths of type *x-y* with an out-shift: we move *p* from *c* towards *a* and we move *q* from *c* towards *b*. If *p* reaches *a* before *q* reaches *b*, then *p* remains at *a* and *q* continues to move towards *b*. Likewise, *p* continues to move towards *a* if *q* reaches *b*. In Phase I, the current shortcut is the best shortcut encountered so far. Phase I ends when the shortcut reaches the end of the backbone, i.e., $pq = ab$, or when a second type of diametral pair appears.

We might switch directly from Phase I to Phase III and skip Phase II when diametral pairs of type *x*-■ and ■-*y* appear simultaneously. For instance, suppose that $T$ is a path and that we move both endpoints of the shortcut with unit speed during Phase I. Then, $x, \bar{x}$ and $\bar{y}, y$ will become diametral pairs of types *x*-● and ●-*y* at the same time.

*Phase II: Shifting Sideways.* The second phase begins when we transit from pair state {*x-y*} to a pair state containing *x-y*. If we transit from {*x-y*} to {*x-y, x*-■}, then we shift towards *x*. If we transit from {*x-y*} to {*x-y,* ■-*y*}, then we shift towards *y*. If we transit from {*x-y*} to {*x-y,* ■-○}, then we branch the search into a shift towards *x* and a shift towards *y*. In the following, we discuss the *x*-shift in Phase II for the pair states {*x-y, x*-■}, {*x-y, x*-■, ■-○}, and {*x-y,* ■-○}. The *y*-shift in Phase II for the pair states {*x-y,* ■-*y*}, {*x-y,* ■-*y,* ■-○}, and {*x-y,* ■-○} is symmetric.

Suppose we reach the pair state {*x-y, x*-■} from {*x-y*}. All diametral paths in $T + pq$ contain the path from *a* to *p*. We move *p* closer to *a*, thereby shrinking the diameter. At the same time, we move *q* with a speed towards *a* that keeps all diametral paths in balance. Thus, we remain in the current pair state until another diametral pair appears. In state {*x-y, x*-■}, the current shortcut is the best shortcut encountered so far.

When we reach the pair state {*x-y,* ■-○} then we move *p* towards *a* and adjust the position of *q* to balance the diametral paths of type *x-pq-y* with those of type ■-*p*-○ and ■-*q*-○. In this state, the diameter shrinks and grows with the length of the shortcut and the best shortcut so far is the shortest shortcut encountered since we entered this state.

Balancing the diametral paths when moving *pq* ensures that we remain in the current pair state until another diametral pair appears. It also restricts our search considerably: we are performing a linear search, since the speed of *q* is determined by the speed of *p*, the path state, and the change in the length of the shortcut, as shown in Section 6.

Phase II ends when *p* reaches *a*, when we transit to a pair state containing *x-y* and ■-■, when we transit from {*x-y,* ■-○} to the final pair state {*x-y,* ■-○, ■-*y*}, or when we transit from {*x-y, x*-■} to the pair state {*x-y, x*-■, ■-*y*} where Phase III begins.

*Phase III: Shifting Outwards.* Phase III starts when we reach {*x-y, x*-■, ■-*y*} from {*x-y, x*-■} or from {*x-y,* ■-*y*}. Since *x-y, x*-■, and ■-*y* block all other movements, we shift outwards balancing *x*-■ and ■-*y*. We immediately transit to {*x*-■, ■-*y*}, since the path from *x* to *y* via the shortcut shrinks faster than the diametral paths connecting *x*-■ and ■-*y*. If we reach Phase III, then the shortest shortcut encountered during Phase III is optimal. Phase III ends when *p* meets *a*, when *q* meets *b*, or when ■-■ or ■-○ appears.

*5.2. Optimality*

We argue that the shortcut produced by the above algorithm is indeed optimal, using invariants for each pair state that follow from the blocking lemma (Lemma 10).

Moreover, we show that one endpoint of the shortcut remains on the path from $a$ to $c$ while the other endpoint remains on the path from $c$ to $b$ along the backbone.

While the algorithm is in Phase I, there is an optimal shortcut $p^*q^*$ that we reach from the current shortcut $pq$ by shifting upwards, by shifting towards $x$, by shifting towards $y$, or by remaining stationary. This invariant holds because the diametral pairs of type $x$-$y$ block any inward shift, i.e., we have $\mathrm{diam}(T + pq) \leq \mathrm{diam}(T + p'q')$ for any shortcut $p'q'$ that we reach with an inward shift from $pq$, due to Lemma 10. Therefore, if Phase I ends with $pq = ab$, then $ab$ is an optimal shortcut for $T$.

While the algorithm is in pair state $\{x\text{-}y, x\text{-}\blacksquare\}$ of Phase II, there is an optimal shortcut $p^*q^*$ that we reach from the current shortcut $pq$ by shifting towards $x$, by shifting outwards, or by remaining stationary. This invariant holds because the diametral pairs of type $x$-$y$ and $x$-$\blacksquare$ in $T + pq$ block any inward shift and any shift towards $y$, i.e., we have $\mathrm{diam}(T + pq) \leq \mathrm{diam}(T + p'q')$ for any shortcut $p'q'$ that we reach with an inward shift or a shift towards $y$ from $pq$, due to Lemma 10. Therefore, if Phase II concludes with $p = a$ in pair state $\{x\text{-}y, x\text{-}\blacksquare\}$, then the current shortcut is optimal, because when $p = a$, every shift towards $x$ is also an inwards shift and, thus, blocked by $x$-$y$, and every outwards shift is also a shift towards $y$ and, thus, blocked by $x$-$\blacksquare$.

While the algorithm is in pair state $\{x\text{-}y, \blacksquare\text{-}\circ\}$ of Phase II, there is an optimal shortcut $p^*q^*$ that we reach from the current shortcut $pq$ by shifting towards $x$, by shifting towards $y$, or by remaining stationary. This invariant holds because the diametral pairs of type $x$-$y$ and $\blacksquare$-$\circ$ in $T + pq$ block any in-shift and any out-shift, i.e., we have $\mathrm{diam}(T + pq) \leq \mathrm{diam}(T + p'q')$ for any shortcut $p'q'$ that we reach with an inward or outward shift from $pq$, due to Lemma 10. We cannot miss an optimal shortcut:

**Invariant 1** *Suppose there is an optimal shortcut $p^*q^*$ for $T$ in the direction of an $x$-shift when the algorithm transits to the pair state $\{x\text{-}y, \blacksquare\text{-}\circ\}$ for the first time. While performing an $x$-shift in pair state $\{x\text{-}y, \blacksquare\text{-}\circ\}$ of Phase II, we have already encountered an optimal shortcut or we can reach $p^*q^*$ from the current shortcut $pq$ with an $x$-shift.*

PROOF. Suppose we perform an $x$-shift from $pq$ while balancing the diametral paths $x$-$y$ and $\blacksquare$-$\circ$ until the pair state changes at some position $p'q'$. If the movement from $p'q'$ to $p^*q^*$ is an $x$-shift, then so is the movement from $pq$ to $p^*q^*$, by transitivity.

Suppose $pq$ to $p^*q^*$ is an $x$-shift while $p'q'$ to $p^*q^*$ is not an $x$-shift. We argue that we encounter an optimal shortcut while shifting from $pq$ to $p'q'$. Let $p''q''$ be the last position during the shift from $pq$ to $p'q'$ where the movement from $p''q''$ to $p^*q^*$ is an $x$-shift. Then $T + p''q''$ is in pair state $\{x\text{-}y, \blacksquare\text{-}\circ\}$ and $p'' = p^*$ or $q'' = q^*$. If $p'' = p^*$, then $p''q''$ to $p^*q^*$ is an inward shift towards $x$. Since $x$-$y$ blocks any in-shift, $\mathrm{diam}(T+p''q'') \leq \mathrm{diam}(T+p^*q^*)$. Likewise, if $q'' = q^*$, then $p''q''$ to $p^*q^*$ is an outward shift towards $x$. Since $\blacksquare$-$\circ$ blocks any out-shift, $\mathrm{diam}(T + p''q'') \leq \mathrm{diam}(T + p^*q^*)$. In both cases, $p''q''$ is optimal, as $p^*q^*$ is optimal. Therefore, we have encountered an optimal shortcut when the movement from $pq$ to $p^*q^*$ is no longer an $x$-shift. □

If Phase II ends in $\{x\text{-}y, \blacksquare\text{-}\circ\}$ with $p = a$ or in the final state $\{x\text{-}y, \blacksquare\text{-}\circ, \blacksquare\text{-}y\}$, then all directions are blocked and, by the invariant, we have encountered an optimal shortcut.

When the algorithm enters Phase III with a transition to $\{x\text{-}y, x\text{-}\blacksquare, \blacksquare\text{-}y\}$, then there is an optimal shortcut $p^*q^*$ that we reach from the current shortcut $pq$ by shifting outwards or by remaining stationary. This is because all movements, except for the out-shift, are blocked by the diametral pairs of types $x$-$y$, $x$-■, and ■-$y$. As we perform an out-shift in the pair state $\{x$-■, ■-$y\}$ of Phase III, we have already encountered an optimal shortcut or the optimal shortcut $p^*q^*$ can still be reached with an out-shift from the current shortcut.

Therefore, we have encountered an optimal shortcut when Phase III ends in the final state $\{x$-■, ■-$y$, ■-■$\}$, or in $\{x$-■, ■-$y$, ■-○$\}$, or when $p = a$ or $q = b$ in the state $\{x$-■, ■-$y\}$.

Finally, we argue that the endpoints of the shortcut remain on their respective sub-paths of the backbone, i.e., the absolute center $c$ remains on the path from $p$ to $q$.

**Invariant 2** *At any moment during the course of the continuous algorithm, the point $p$ lies on the path from $a$ to $c$ in $T$ and the point $q$ lies on the path from $c$ to $b$ in $T$.*

PROOF. The invariant holds at the start of Phase I, since we begin by setting $pq = cc$. Throughout Phase I, we perform an out-shift that upholds the invariant: both $p$ and $q$ move away from $c$ and neither does $p$ move past $a$ nor does $q$ move past $b$.

For Phase II, we prove that $q$ cannot pass through $c$ during an $x$-shift and, symmetrically, that $p$ cannot pass through $c$ during $y$-shift. Hence, the invariant holds throughout Phase II, since the algorithm terminates when $p$ reaches $a$ or when $q$ reaches $b$.

Assume, for a contradiction, that $q$ reaches $c$ when performing an $x$-shift in Phase II. Without loss of generality, let this be the first time $q$ reaches $c$ during Phase II. Hence, the invariant holds until now and, thus, $p$ lies on the path from $a$ to $c$. Since we are in Phase II, there are diametral pairs of type $x$-$y$ in $T + pq$ as well as diametral pairs of type $x$-■ or ■-○. We distinguish two cases depending on the pair state of the augmented tree $T + pc$. In each case, we derive a contradiction, i.e., $q$ could never have reached $c$.

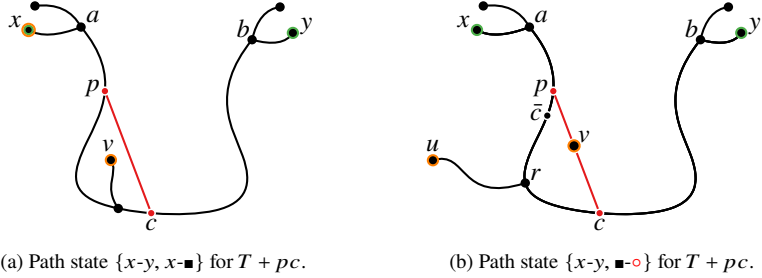

(a) Path state $\{x$-$y$, $x$-■$\}$ for $T + pc$.  (b) Path state $\{x$-$y$, ■-○$\}$ for $T + pc$.

Figure 23: Two impossible cases when $T + pc$ has diametral pairs of types (a) $x$-$y$ and $x$-■ or (b) $x$-$y$ and ■-○.

1. Suppose $T + pc$ has diametral pairs of type $x$-$y$ and $x$-■, as illustrated in Figure 23a. Let $x, v$ be a diametral pair of type $x$-■ in $T + pc$. Since $v$ does not lie in a primary $\mathcal{B}$-sub-tree of $T$, we have $d_T(c, v) < d_T(c, y)$. Since $T + pc$ has other diametral pairs besides those of type $x$-$y$, the shortcut $pc$ must be useful for $(x, y)$, i.e., $d_T(x, p) + |pc| + d_T(c, y) = d_{T+pc}(x, y)$. This leads to the contradiction $\mathrm{diam}(T + pc) < \mathrm{diam}(T + pc)$, since

$$\mathrm{diam}(T + pc) = d_{T+pc}(x, v) \qquad (x, v \text{ is diametral in } T + pc)$$
$$\leq d_T(x, p) + |pc| + d_T(c, v)$$

$$
\begin{aligned}
&< d_T(x, p) + |pc| + d_T(c, y) && (d_T(c, v) < d_T(c, y)) \\
&= d_{T+pc}(x, y) && (pc \text{ is useful for } (x, y)) \\
&= \text{diam}(T + pc) \ . && (x, y \text{ is diametral in } T + pc)
\end{aligned}
$$

2. Suppose $T + pc$ has diametral pairs of type $x$-$y$ and ■-○, as illustrated in Figure 23b. Let $u, v$ be a diametral pair of type ■-○ in $T + pc$ with $v \notin T$. Then $v \in pc$ with $p \neq v \neq c$ and $u$ lies in a secondary $\mathcal{B}$-sub-tree with root $r$ attached to the path from $p$ to $c$ along $\mathcal{B}$. Let $\bar{c}$ be the farthest point from $c$ on the simple cycle $C(p, c)$ in $T + pc$. This leads to the contradiction $\text{diam}(T + pc) < \text{diam}(T + pc)$, since

$$
\begin{aligned}
\text{diam}(T + pc) &= d_{T+pc}(u, v) && (u, v \text{ is diametral in } T + pc) \\
&= d_T(u, r) + d_{T+pc}(r, v) && (u, v \text{ is of type ■-○}) \\
&\leq d_T(u, c) + d_{T+pc}(r, v) && (r \text{ is on the path from } p \text{ to } c) \\
&< d_T(y, c) + d_{T+pc}(r, v) && (u \text{ is in a secondary } \mathcal{B}\text{-sub-tree}) \\
&= d_T(y, c) + (|pc| + d_T(p, c))/2 \\
&&& (r \text{ and } v \text{ are antipodal along } C(p, c)) \\
&= d_T(y, c) + d_{T+pc}(c, \bar{c}) && (c \text{ and } \bar{c} \text{ are antipodal along } C(p, c)) \\
&= d_{T+pc}(y, \bar{c}) && (c \text{ is on any path from } y \text{ to } \bar{c} \text{ in } T + pc) \\
&\leq \text{diam}(T + pc) \ .
\end{aligned}
$$

Therefore, during Phase II, $q$ remains on the path from $c$ to $b$ and, $p$ remains on the path from $a$ to $c$. This invariant holds for the entire algorithm, since the out-shift of Phase III cannot move $p$ or $q$ through $c$ and it ends when $p$ reaches $a$ or when $q$ reaches $b$. □

If we could implement and run the continuous algorithm, then it would produce an optimal shortcut $pq$ for $T$ where $p$ lies along the path from $a$ to $c$ and $q$ lies along the path from $c$ to $b$. We simulate the continuous algorithm with a discretization.

## 6. Discretization

To discretize the continuous algorithm, we subdivide the continuous motion of the shortcut with events such that we can calculate the next event and the change in the continuous diameter of $T + pq$ between subsequent events. We introduce events when the shortcut meets a vertex, when the path state changes, and when the shortcut begins to shrink or to grow.

### 6.1. Simulating Phase I

In Phase I of the continuous algorithm, we move $p$ and $q$ with unit speed towards $a$ and $b$, respectively. In the discrete algorithm, we process a vertex event whenever $p$ or $q$ would meet a vertex during the continuous movement. We locate the next vertex event by comparing the distance from $p$ and from $q$ to the next vertex along their respective paths towards $a$ and towards $b$. Phase I begins in path state $\{x\text{-}pq\text{-}y, x\text{-}T\text{-}y\}$ with $pq = cc$; the diametral paths of type $x\text{-}T\text{-}y$ disappear when the shortcut becomes useful for $T$.

During Phase I, the continuous diameter decreases or remains constant. Therefore, it is sufficient for the discrete algorithm to determine where Phase I of the continuous algorithm ends. At the end of Phase I, we have either reached the end of the backbone, i.e., $pq = ab$, or a diametral pair of type $x$-■, ■-$y$, or ■-○ has appeared alongside the diametral pairs of type $x$-$y$. We may ignore diametral pairs of type ■-■ as they appear together with $x$-■ and ■-$y$ when $x$-$y$ is diametral, as shown in Theorem 12.

We detect changes in the path state by monitoring the candidates for each type of diametral path—except for those connecting diametral pairs of type ■-■—as follows.

$x$-$pq$-$y$ The path $x$-$pq$-$y$ has length $d_{T+pq}(x, y) = d_T(x, p) + |pq| + d_T(q, y)$. Between two subsequent events, $p$ and $q$ remain on their respective containing edge. If the edges of $T$ are straight-line segments, then we can express the positions of $p$ and $q$ as an algebraic function of time with constant degree. Therefore, we can also express $d_{T+pq}(x, y)$ as an algebraic function of constant degree.

$x$-$pq$-▲, ▲-$pq$-$y$ If the path $x$-$pq$-$s_i$, for $i = 1, 2, \ldots, k$ becomes a diametral path of type $x$-$pq$-▲, then $pq$ is useful or indifferent for $(x, s_i)$, i.e., the leaf $s_i$ belongs to a secondary $\mathcal{B}$-sub-tree $S_i$ that is attached to the path from $\bar{p}$ to $b$ in $T$. If $x$-$pq$-$s_i$ becomes diametral in Phase I or II, then $x$-$pq$-$y$ is also diametral and $S_i$ is attached to the path from $q$ to $\bar{p}$, as otherwise $y$ would be farther from $x$ than $s_i$ in $T + pq$. Therefore, for any $i = 1, 2, \ldots, k$, the paths $x$-$pq$-$y$ and $x$-$pq$-$s_i$ are diametral in $T + pq$ if and only if $pq$ is useful or indifferent for $(x, s_i)$ and $d_T(q, y) = d_T(q, s_i)$, i.e., $q$ lies midway along the path from $s_i$ to $y$.

We introduce new events at the points $q_i \in T$ with $d_T(q_i, y) = d_T(q_i, s_i)$, for $i = 1, 2, \ldots, k$, whenever $q_i$ lies on the path from $c$ to $b$. To locate the points $q_1, q_2, \ldots, q_k$, we first sort the distances $d_T(y, s_1), d_T(y, s_2), \ldots, d_T(y, s_k)$ and then traverse the path from $b$ to $c$ placing each $q_i$ at the appropriate distance from $y$. Thus, locating $q_1, q_2, \ldots, q_k$ takes $O(n + k \log k) = O(n \log n)$ time.

When $q$ reaches $q_i$, for some $i = 1, 2, \ldots, k$, during Phase I we test whether $pq$ is useful or indifferent for $(x, s_i)$. If $pq$ is useful or indifferent for $(x, s_i)$, then $x$-$pq$-$s_i$ becomes diametral and Phase I ends. Otherwise, $pq$ is useless for $(x, s_i)$ and $x$-$pq$-$s_i$ cannot become diametral during Phase I. Processing this event takes constant time, because the shortcut $pq$ is useful or indifferent for $(x, s_i)$ if and only if the root $r_i$ of $S_i$ lies on the path from $\bar{p}$ to $b$ in $T + pq$, which we can check in constant time by comparing $d_T(a, r_i)$ with $d_T(a, \bar{p})$, since

$$d_T(a, \bar{p}) = d_T(a, p) + d_T(p, \bar{p}) = d_T(a, p) + \frac{|pq| + d_T(p, q)}{2} \ .$$

$x$-$T$-▲, ▲-$T$-$y$ The length of the path $x$-$T$-$s_i$ for some $i = 1, 2, \ldots k$ does not change with $pq$; what does change is whether $x$-$T$-$s_i$ is a *shortest* path in $T + pq$ or not: the shortcut $pq$ is useful for $(x, s_i)$ if and only if $\bar{p}$ lies in the interior of the path from $a$ to $r_i$ in $T$. Otherwise, $pq$ is indifferent for $\{x, s_i\}$, i.e., $d_{T+pq}(x, s_i) = d_T(x, s_i)$.

The path $x$-$T$-$s_i$ becomes a diametral path of type $x$-$T$-▲ when the continuous diameter of $T + pq$ decreases to $d_T(x, s_i)$ while the shortcut $pq$ is indifferent for $\{x, s_i\}$. To detect this, we sort the values $d_T(x, s_1), d_T(x, s_2), \ldots, d_T(x, s_k)$ and we introduce events when the current continuous diameter reaches any of these values. This leads to $O(n \log n)$ additional preprocessing time and $O(k) = O(n)$ additional events, since the continuous diameter is decreasing during Phase I. For each additional event, we compare $d_T(a, r_i)$ with $d_T(a, \bar{p})$ to decide, in constant time, whether $pq$ is indifferent for $\{x, s_i\}$.

*x-pq-●, ●-pq-y, x-T-●, ●-T-y, ▲-p-○, ▲-q-○, ▲-pq-●, ▲-T-●* We turn to diametral paths in $T + pq$ that connect a leaf $l$ of $T$ with the farthest point from $l$ along $C(p, q)$. Any candidate for a diametral path of this kind has length $h + \frac{1}{2}(d_T(p, q) + |pq|)$, where $h$ is the height of a tallest sub-tree attached to the cycle $C(p, q)$ in $T + pq$.

At the beginning of Phase I, the sub-trees containing $x$ and $y$ are the tallest sub-trees attached to $C(p, q)$, i.e., $h = d_T(x, p) = d_T(q, y)$. As Phase I progresses, these sub-trees shrink as $p$ moves to $a$ and $q$ moves to $b$. For any secondary $\mathcal{B}$-sub-tree $S_i$ attached to the path from $a$ to $p$, we have $d_T(r_i, s_i) \leq d_T(p, s_i) < d_T(p, x)$, and for every secondary $\mathcal{B}$-sub-tree $S_j$ attached to the path from $q$ to $b$, we have $d_T(r_j, s_j) \leq d_T(q, s_j) < d_T(q, y)$. Thus, a secondary $\mathcal{B}$-sub-tree $S_i$ may only become a tallest sub-tree if its root $r_i$ lies on the path from $p$ to $q$ and $h = \max\{d_T(x, p), d_T(r_1, s_1), d_T(r_2, s_2), \ldots, d_T(r_k, s_k), d_T(q, y)\}$.

We compute $\hat{h} = \max\{d_T(r_1, s_1), d_T(r_2, s_2), \ldots, d_T(r_k, s_k)\}$ as part of our preprocessing. We detect diametral paths of type *x-pq-●, ●-pq-y, x-T-●, ●-T-y, ▲-p-○, ▲-q-○, ▲-pq-●*, and *▲-T-●* by comparing the current continuous diameter with

$$\max\{d_T(x, p), \hat{h}, d_T(q, y)\} + \frac{1}{2}(d_T(p, q) + |pq|) \ .$$

This takes constant additional time per event.

Thus, we can simulate Phase I of the continuous algorithm in $O(n \log n)$ time.

### 6.2. Simulating Phase II

We describe the discretization of Phase II for a shift towards $x$, where the continuous algorithm balances the diametral path *x-pq-y* and a diametral path with endpoints of type *x-■* or *■-○*. Suppose $p$ moves with unit speed towards $a$. The following lemma specifies the speed at which $q$ should move towards $c$ to balance the diametral paths.

**Lemma 13** *Let $pq$ and $p'q'$ be two shortcuts for a geometric tree $T$ where we reach $p'q'$ from $pq$ with an x-shift. If $T + pq$ and $T + p'q'$ are in the same path state, then we can express the distance from $q$ to $q'$ and the change in diameter as stated in Table 1.*

| Path State | $d_T(q, q')$ | $\text{diam}(T + pq) - \text{diam}(T + p'q')$ |
|---|---|---|
| $\{x\text{-}pq\text{-}y, x\text{-}pq\text{-}▲\}$ | 0 | $d_T(p', p) + |pq| - |p'q'|$ |
| $\{x\text{-}pq\text{-}y, x\text{-}pq\text{-}●, x\text{-}T\text{-}●\}$ | $\frac{1}{3}(d_T(p, p') + |pq| - |p'q'|)$ | $\frac{2}{3}(d_T(p, p') + |pq| - |p'q'|)$ |
| $\{x\text{-}pq\text{-}y, x\text{-}T\text{-}▲\}$ | $d_T(p, p') + |pq| - |p'q'|$ | 0 |
| $\{x\text{-}pq\text{-}y, ▲\text{-}p\text{-}○, ▲\text{-}q\text{-}○\}$ | $d_T(p, p') + \frac{1}{3}(|pq| - |p'q'|)$ | $\frac{2}{3}(|pq| - |p'q'|)$ |

Table 1: The distance between $q$ and $q'$ and the change in diameter when shifting the shortcut towards $x$ from $pq$ to $p'q'$ while maintaining the diametral paths in balance.

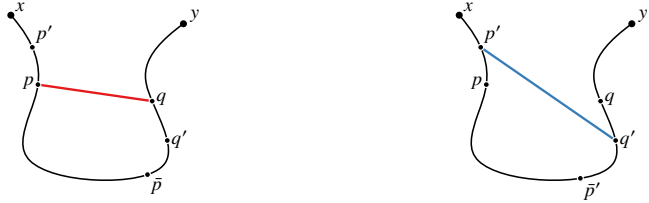

Figure 24: The situation at the start (left) and at the end (right) of an $x$-shift from $pq$ to $p'q'$.

Proof. We show the result for the path state $\{x\text{-}pq\text{-}y, x\text{-}pq\text{-}\bullet, x\text{-}T\text{-}\bullet\}$. During an $x$-shift from $pq$ to $p'q'$, as in Figure 24, the length of the paths of type $x\text{-}pq\text{-}y$ changes by

$$d_{T+pq}(x, y) - d_{T+p'q'}(x, y) = [d_T(x, p') + d_T(p', p) + |pq| + d_T(q, y)]$$
$$- [d_T(x, p') + |p'q'| + d_T(q', q) + d_T(q, y)]$$
$$= d_T(p', p) + |pq| - |p'q'| - d_T(q', q) \ .$$

Let $\bar{p}$ and $\bar{p}'$ be the farthest point from $p$ along $C(p, q)$ and along $C(p', q')$, respectively. Each diametral path of type $x\text{-}pq\text{-}\bullet$ and $x\text{-}T\text{-}\bullet$ connects $x$ with $\bar{p}$ in $T + pq$. Since $p$ and $\bar{p}$ are antipodal along $C(p, q)$, we have $d_{T+pq}(p, \bar{p}) = |C(p, q)|/2 = (|pq| + d_T(p, q))/2$. Therefore, the length of the paths of type $x\text{-}pq\text{-}\bullet$ and $x\text{-}T\text{-}\bullet$ in $T + pq$ is

$$d_{T+pq}(x, \bar{p}) = d_T(x, p) + d_{T+pq}(p, \bar{p}) = d_T(x, p) + \frac{|pq| + d_T(p, q)}{2} \ ,$$

Likewise, the lengths of the paths of type $x\text{-}pq\text{-}\bullet$ and $x\text{-}T\text{-}\bullet$ in $T + p'q'$ is $d_{T+p'q'}(x, \bar{p}') = d_T(x, p') + (|p'q'| + d_T(p', q'))/2$. During an $x$-shift from $pq$ to $p'q'$, the length of the paths of type $x\text{-}pq\text{-}\bullet$ and $x\text{-}T\text{-}\bullet$ changes by

$$d_{T+pq}(x, \bar{p}) - d_{T+p'q'}(x, \bar{p}') = \left[d_T(x, p') + d_T(p', p) + \frac{|pq| + d_T(q, q') + d_T(q', p)}{2}\right]$$
$$- \left[d_T(x, p') + \frac{|p'q'| + d_T(q', p) + d_T(p, p')}{2}\right]$$
$$= \frac{d_T(p, p') + |pq| - |p'q'| + d_T(q, q')}{2} \ .$$

Since $x\text{-}pq\text{-}y$, $x\text{-}pq\text{-}\bullet$, and $x\text{-}T\text{-}\bullet$ remain diametral during the $x$-shift from $pq$ to $p'q'$, their lengths change by the same amount, i.e., $d_{T+pq}(x, y) - d_{T+p'q'}(x, y) = d_{T+pq}(x, \bar{p}) - d_{T+p'q'}(x, \bar{p}')$ and, thus, $d_T(q, q') = \frac{1}{3}(d_T(p, p') + |pq| - |p'q'|)$, since

$$3d_T(q, q') = 3d_T(q, q') + 2 \cdot 0$$
$$= 3d_T(q, q') + 2\left[d_{T+pq}(x, y) - d_{T+p'q'}(x, y)\right]$$
$$- 2\left[d_{T+pq}(x, \bar{p}) - d_{T+p'q'}(x, \bar{p}')\right]$$
$$= 3d_T(q, q') + [2d_T(p', p) + 2|pq| - 2|p'q'| - 2d_T(q', q)]$$
$$- [d_T(p, p') + |pq| - |p'q'| + d_T(q, q')]$$
$$= d_T(p', p) + |pq| - |p'q'| \ .$$

In this case, the diameter changes by $\frac{2}{3}(d_T(p, p') + |pq| - |p'q'|)$, since

$$\begin{aligned}
\text{diam}(T + pq) - \text{diam}(T + p'q') &= d_{T+pq}(x, y) - d_{T+p'q'}(x, y) \\
&= d_T(p', p) + |pq| - |p'q'| - d_T(q', q) \\
&= d_T(p', p) + |pq| - |p'q'| - \frac{d_T(p, p') + |pq| - |p'q'|}{3} \\
&= \frac{2}{3}[d_T(p, p') + |pq| - |p'q'|] \quad .
\end{aligned}$$

Analogously, we establish the results for the other path states listed in Table 1. $\square$

When $p$ and $q$ traverse a fixed pair of edges while shifting towards $x$, Lemma 13 allows us to compute the next vertex event and how the diameter changes between subsequent events. While simulating Phase II, we encounter $O(n)$ events where the shortcut meets a vertex or where the shortcut begins to shrink or grow, due to the following. The shortcut enters each edge at most once, since $p$ and $q$ never change direction (we have $d_T(p, p') > 0$ and $d_T(q, q') \geq 0$ from Lemma 13). Thus, we encounter $O(n)$ pairs of edges. Moreover, the shortcut changes at most once between growing and shrinking when both endpoints move along a fixed pair of edges.

**Lemma 14** *For a geometric tree with n vertices and straight-line edges, the path state changes $O(n)$ times during Phase II of the continuous algorithm we can locate the next event where the path state changes in $O(\log n)$ time, after $O(n \log n)$ preprocessing time.*

PROOF. In Phase II, the speed of $p$ and $q$ is determined by two or three different types of diametral paths, i.e., by *x-pq-y* and *x-pq-▲*, or by *x-pq-y* and *x-pq-●* and *x-T-●*, or *x-pq-y* and *x-T-▲*, or by *x-pq-y* and *▲-p-○* and *▲-q-○*. For any path state $X$ during Phase II, only the subset $X'$ of $X$ that leads to the least increase in the continuous diameter determines the speed of $p$ and $q$. The other types of paths cease to be diametral instantaneously, i.e., we transit from $X$ to $X'$. Some transitions between the path states in Phase II are impossible. For instance, suppose we perform an *x*-shift from $pq$ to a new position $p'q'$ such that the path state remains $\{x\text{-}pq\text{-}y, ▲\text{-}p\text{-}○, ▲\text{-}q\text{-}○\}$ until we reach $p'q'$ and where the path state changes upon reaching $p'q'$. Then $T + p'q'$ cannot be in path state $\{x\text{-}pq\text{-}y, ▲\text{-}p\text{-}○, ▲\text{-}q\text{-}○, x\text{-}pq\text{-}●, x\text{-}T\text{-}●\}$, due to the following. From Table 1 we know $d_T(q, q') = d_T(p, p') + \frac{1}{3}(|pq| - |p'q'|)$. As $pq$ moves to $p'q'$, the length of the paths *x-pq-●* and *x-T-●* changes by $d_T(p', p) + \frac{2}{3}(|pq| - |p'q'|)$, since

$$\begin{aligned}
d_{T+pq}(x, \bar{p}) - d_{T+p'q'}(x, \bar{p}') &= \frac{d_T(p', p) + |pq| - |p'q'| + d_T(q, q')}{2} \\
&= \frac{d_T(p', p) + |pq| - |p'q'| + d_T(p, p') + \frac{1}{3}(|pq| - |p'q'|)}{2} \\
&= d_T(p', p) + \frac{2}{3}(|pq| - |p'q'|) \quad .
\end{aligned}$$

Since $d_T(p, p') > 0$, the paths of type *x-pq-●* and *x-T-●* shrink at a faster rate than the diametral paths of type *x-pq-y*, *▲-p-○*, and *▲-q-○*. Therefore, *x-pq-●* and *x-T-●* cannot become diametral when the shortcut reaches $p'q'$. Figure 25 illustrates the transitions between the path states that may occur during Phase II.

We bound the number of visits to the path state $\{x\text{-}pq\text{-}y, x\text{-}pq\text{-}\blacktriangle\}$ by $k$, where $k = O(n)$ is the number of secondary $\mathcal{B}$-sub-trees of $T$. When we are in path state $\{x\text{-}pq\text{-}y, x\text{-}pq\text{-}s_j\}$ for some $j = 1, 2, \ldots, k$, then $q$ lies midway along the path from $y$ to $s_j$, i.e., $d_T(q, y) = d_T(q, s_j)$. When we transit from $\{x\text{-}pq\text{-}y, x\text{-}pq\text{-}s_j\}$ to any other path state, $q$ will begin to move with non-zero speed towards $a$. Hence, $x, s_j$ ceases to be diametral and cannot become diametral again during Phase II. Therefore, we take at most $k$ green transitions in Figure 25.

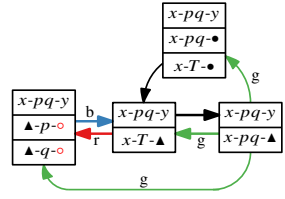

Figure 25: The transitions between the path states during Phase II. Transitory states, such as $\{x\text{-}pq\text{-}y, x\text{-}pq\text{-}\blacktriangle, x\text{-}pq\text{-}\bullet, x\text{-}T\text{-}\bullet\}$ have been omitted. The red transition (r) is possible when the shortcut grows; the blue transition (b) is possible when the shortcut shrinks. When we take any green transition (g) from $\{x\text{-}pq\text{-}y, x\text{-}pq\text{-}s_j\}$ for $j = 1, 2, \ldots, k$ then $x, s_j$ ceases to be diametral and cannot become diametral again.

We bound the number of visits to the pair state $\{x\text{-}pq\text{-}y, \blacktriangle\text{-}p\text{-}\circ, \blacktriangle\text{-}q\text{-}\circ\}$ by $O(n)$. Once the green transitions are exhausted, we can only enter $\{x\text{-}pq\text{-}y, \blacktriangle\text{-}p\text{-}\circ, \blacktriangle\text{-}q\text{-}\circ\}$ with the red transition from $\{x\text{-}pq\text{-}y, x\text{-}T\text{-}\blacktriangle\}$. This transition is only possible when the shortcut is growing. Therefore, we pay $O(n)$ visits to $\{x\text{-}pq\text{-}y, \blacktriangle\text{-}p\text{-}\circ, \blacktriangle\text{-}q\text{-}\circ\}$, since the shortcut switches at most $2n$ times between shrinking and growing during an $x$-shift.

After $O(n)$ visits to the path states $\{x\text{-}pq\text{-}y, x\text{-}pq\text{-}\blacktriangle\}$ and $\{x\text{-}pq\text{-}y, \blacktriangle\text{-}p\text{-}\circ, \blacktriangle\text{-}q\text{-}\circ\}$, we can no longer take any red or green transitions in Figure 25. We encounter $O(n)$ path state changes during Phase II, as the remaining transitions form an acyclic digraph. □

Thus, we can simulate Phase II in $O(n \log n)$ time, since Phase II consists of $O(n)$ events that we process in $O(\log n)$ time after $O(n \log n)$ preparation time by monitoring the diametral paths of the augmented tree $T + pq$ in the same fashion as in Phase I.

*6.3. Simulating Phase III*

In Phase III, the continuous algorithm balances a diametral path with endpoints of type $x$-■ and a diametral path with endpoints of type ■-$y$. This leads to the following.

**Lemma 15** *Let $pq$ and $p'q'$ be two shortcuts for a geometric tree $T$ where we reach $p'q'$ from $pq$ with an out-shift. If $T + pq$ and $T + p'q'$ are in the same path state, then we can express the distance of $q$ and $q'$ and the change in diameter as stated in Table 2.*

| $x$-■ | ■-$y$ | $d_T(q,q')$ | $\mathrm{diam}(T+pq) - \mathrm{diam}(T+p'q')$ |
| --- | --- | --- | --- |
| $x\text{-}pq\text{-}\blacktriangle$ | $\blacktriangle\text{-}pq\text{-}y$ | $d_T(p,p')$ | $\lvert pq \rvert - \lvert p'q' \rvert$ |
| $x\text{-}pq\text{-}\blacktriangle$ | $\bullet\text{-}y$ | $d_T(p,p') + \frac{1}{3}(\lvert pq \rvert - \lvert p'q' \rvert)$ | $\frac{2}{3}(\lvert pq \rvert - \lvert p'q' \rvert)$ |
| $x\text{-}pq\text{-}\blacktriangle$ | $\blacktriangle\text{-}T\text{-}y$ | $d_T(p,p') + \lvert pq \rvert - \lvert p'q' \rvert$ | 0 |
| $x\text{-}\bullet$ | $\blacktriangle\text{-}pq\text{-}y$ | $d_T(p,p') - \frac{1}{3}(\lvert pq \rvert - \lvert p'q' \rvert)$ | $\frac{2}{3}(\lvert pq \rvert - \lvert p'q' \rvert)$ |
| $x\text{-}\bullet$ | $\bullet\text{-}y$ | $d_T(p,p')$ | $\frac{1}{2}(\lvert pq \rvert - \lvert p'q' \rvert)$ |
| $x\text{-}\bullet$ | $\blacktriangle\text{-}T\text{-}y$ | $d_T(p,p') + \lvert pq \rvert - \lvert p'q' \rvert$ | 0 |
| $x\text{-}T\text{-}\blacktriangle$ | $\blacktriangle\text{-}pq\text{-}y$ | $d_T(p,p') - (\lvert pq \rvert - \lvert p'q' \rvert)$ | 0 |
| $x\text{-}T\text{-}\blacktriangle$ | $\bullet\text{-}y$ | $d_T(p,p') - (\lvert pq \rvert - \lvert p'q' \rvert)$ | 0 |
| $x\text{-}T\text{-}\blacktriangle$ | $\blacktriangle\text{-}T\text{-}y$ | $d_T(p,p')$ | 0 |

Table 2: The distance between $q$ and $q'$ with the change in diameter during an out-shift from $pq$ to $p'q'$ while balancing the diametral paths. Here, $x\text{-}\bullet$ stands for $x\text{-}pq\text{-}\bullet$ and $x\text{-}T\text{-}\bullet$; and $\bullet\text{-}y$ stands for $\bullet\text{-}pq\text{-}y$ and $\bullet\text{-}T\text{-}y$.

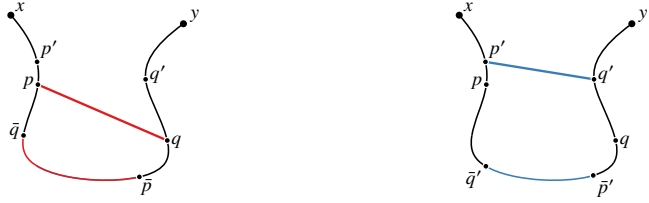

Figure 26: The situation at the start (left) and at the end (right) of an out-shift from $pq$ to $p'q'$.

PROOF. We establish these results using the same approach as in Lemma 13. As $pq$ moves to $p'q'$, as in Figure 26, the potential diametral paths change as follows. For $r \in \{p, q\}$, $\bar{r}$ and $\bar{r}'$ are the farthest point from $r$ along $C(p, q)$ and $C(p', q')$, respectively.

$x$-$pq$-▲ : $\quad d_{T+pq}(x, s_i) - d_{T+p'q'}(x, s_i) = d_T(p', p) + |pq| - |p'q'| - d_T(q', q)$

$x$-$pq$-●, $x$-$T$-● : $\quad d_{T+pq}(x, \bar{x}) - d_{T+p'q'}(x, \bar{x}') = \dfrac{d_T(p', p) + |pq| - |p'q'| - d_T(q', q)}{2}$

$x$-$T$-▲ : $\quad d_{T+pq}(x, s_j) - d_{T+p'q'}(x, s_j) = 0$

▲-$pq$-$y$ : $\quad d_{T+pq}(s_l, y) - d_{T+p'q'}(s_l, y) = d_T(q', q) + |pq| - |p'q'| - d_T(p', p)$

●-$pq$-$y$, ●-$T$-$y$ : $\quad d_{T+pq}(y, \bar{y}) - d_{T+p'q'}(y, \bar{y}') = \dfrac{d_T(q', q) + |pq| - |p'q'| - d_T(p', p)}{2}$

▲-$T$-$y$ : $\quad d_{T+pq}(s_o, y) - d_{T+p'q'}(s_o, y) = 0$

These changes equate for paths that remain diametral during the outwards shift. For instance, if the paths $x$-$pq$-●, $x$-$T$-●, and ▲-$pq$-$y$ remain diametral as $pq$ moves to $p'q'$, then $d_{T+pq}(x, \bar{x}) - d_{T+p'q'}(x, \bar{x}') = \text{diam}(T + pq) - \text{diam}(T + p'q') = d_{T+pq}(s_l, y) - d_{T+p'q'}(s_l, y)$ and, thus, $d_T(p, p') = d_T(q, q') + \frac{1}{3}(|pq| - |p'q'|)$, since

$$\begin{aligned}
3d_T(p, p') &= 3d_T(p, p') + 2 \cdot 0 \\
&= 3d_T(p, p') + 2\left[d_{T+pq}(s_l, y) - d_{T+p'q'}(s_l, y)\right] \\
&\quad - 2\left[d_{T+pq}(x, \bar{x}) - d_{T+p'q'}(x, \bar{x}')\right] \\
&= 3d_T(p, p') + 2[d_T(q', q) + |pq| - |p'q'| - d_T(p', p)] \\
&\quad - [d_T(p', p) + |pq| - |p'q'| - d_T(q', q)] \\
&= 3d_T(q, q') + |pq| - |p'q'| \ .
\end{aligned}$$

In this case, the diameter changes by $\frac{2}{3}(|pq| - |p'q'|)$, since

$$\begin{aligned}
\text{diam}(T + pq) - \text{diam}(T + p'q') &= d_{T+pq}(s_l, y) - d_{T+p'q'}(s_l, y) \\
&= d_T(q', q) + |pq| - |p'q'| - d_T(p', p) \\
&= d_T(q', q) + |pq| - |p'q'| - d_T(q, q') - \dfrac{|pq| - |p'q'|}{3} \\
&= \dfrac{2}{3}(|pq| - |p'q'|) \ .
\end{aligned}$$

In the same manner, we express $d_T(q, q')$ and the change in diameter in terms of $d_T(p, p')$ and $|pq| - |p'q'|$ for the remaining path states listed in Table 2. □

Similar to Phase I and II, there are $O(n)$ events where the shortcut meets a vertex or starts to shrink or grow. Even though we can rule out certain transitions between path states in Phase III, the path state might change $\Omega(n^2)$ times.

**Lemma 16** *For every number $l \in \mathbb{N}$, there exists a geometric tree $T_l$ with $n = 8l + 5$ vertices where at least $3l^2 - 2l = \Omega(n^2)$ path state changes occur in Phase III.*

PROOF (SKETCH). We construct $T_l$ with the following objectives in mind.

1. The geometric tree $T_l$ consist of a backbone path $\mathcal{B}$ from $x$ to $y$ with $2l$ pendant edges $r_1 s_1, r_2 s_2, \ldots, r_{2l} s_{2l}$, where $r_i \in \mathcal{B}$ and $s_i \notin \mathcal{B}$ for each $i = 1, 2, \ldots, 2l$.
2. During Phase III, the shortcut grows and shrinks at least $l$ times.
3. Every time the shortcut $pq$ grows or shrinks, it changes between being useful and useless for the $2l$ pairs $(x, s_{l+1}), (x, s_{l+2}), \ldots (x, s_{2l})$ and $(s_1, y), (s_2, y), \ldots (s_l, y)$. These changes occur at $l$ distinct events. At the $j$-th event, for $j = 1, 2, \ldots, l$, the shortcut switches between being useful and useless for $(x, s_{l+j})$ and $(s_j, y)$.
4. For $j = 1, 2, \ldots, l$, the pairs $(x, s_{l+j})$ and $(s_j, y)$ are the only diametral pairs of $T + pq$ when the shortcut switches between being useful and useless for them.

Figure 27 illustrates our construction for $l = 4$. When the shortcut grows, $\bar{q}$ passes through $r_l, r_{l-1}, \ldots, r_1$ in this order while $\bar{p}$ passes through $r_{l+1}, r_{l+2}, \ldots, r_{2l}$ with $\bar{q}$ reaching $r_j$ at the same time as $\bar{p}$ reaches $r_{2l-j+1}$ for $j = 1, 2, \ldots, l$. By choosing sufficiently short lengths for the edges $r_i s_i$ and by spacing them out sufficiently far apart, we ensure that $x$-$pq$-$s_{l+j}$ and $s_{l-j+1}$-$pq$-$y$ are diametral immediately before the moment when $\bar{q}$ reaches $r_j$ and $\bar{p}$ reaches $s_{l+j}$ and that $x$-$T$-$s_{l+j}$ and $s_{l-j+1}$-$T$-$y$ are diametral shortly after the moment when $\bar{q}$ passed $r_{l-j+1}$ and $\bar{p}$ passed $s_{l+j}$. Furthermore, we achieve that the pairs $(x, \bar{p})$ and $(\bar{q}, y)$ become diametral when $\bar{q}$ and $\bar{p}$ move from $r_{l-j+1}$ to $r_{l-j}$ and from $r_{l+j}$ to $r_{l+j+1}$, respectively. By balancing the lengths of the stretches where the shortcut grows and the lengths of the stretches where the shortcut shrinks, we ensure that the above happens in reverse when the shortcut shrinks.

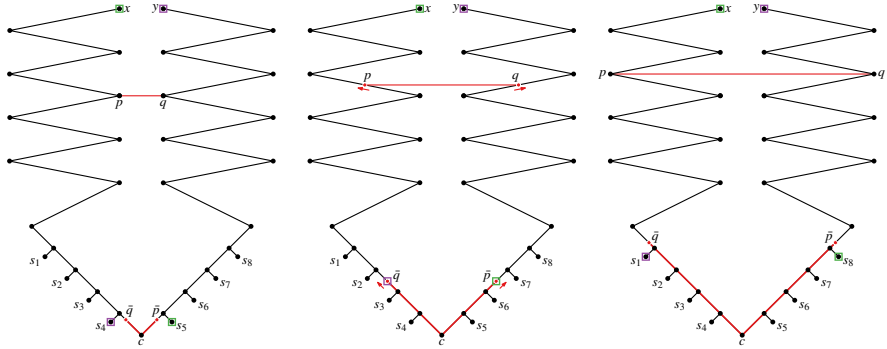

Figure 27: The tree $T_4$ with three positions for the shortcut $pq$ where the shortcut grows from left to right. The diametral pairs of $T + pq$ are indicates with squares of matching color. The sets of diametral paths of $T + pq$ are $\{x$-$pq$-$s_5, s_4$-$pq$-$y\}$, $\{x$-$pq$-$\bar{p}, x$-$T$-$\bar{p}, \bar{q}$-$pq$-$y, \bar{q}$-$T$-$y\}$, and $\{x$-$T$-$s_8, s_1$-$T$-$y\}$ from left to right.

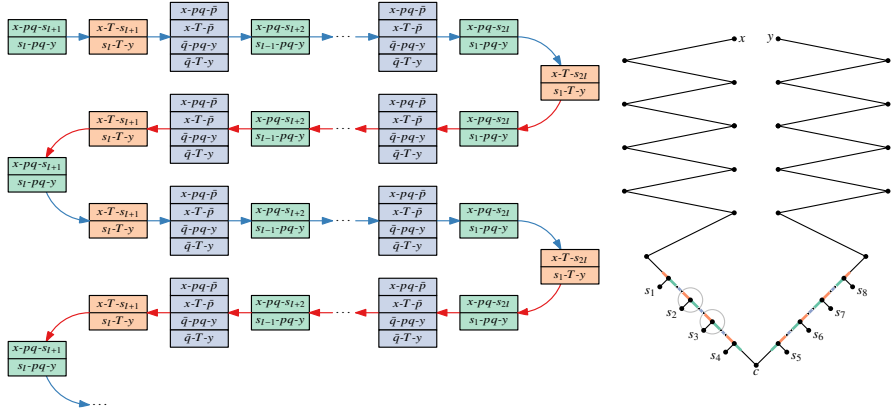

Figure 28: On the left, we represent the changes in the set of diametral paths for the tree $T_l$ during Phase III. Blue transitions occur while the shortcut is growing and red transitions occur while the shortcut is shrinking. On the right, we colour the sections of $T_l$ that are swept by $\bar{q}$ or $\bar{p}$ according to the corresponding path state.

Figure 28 illustrates the path state transitions for $T_l$ during Phase III. We count three path states for each of the $l$ moments when $\bar{q}$ and $\bar{p}$ pass through $r_j$ and $r_{2l-j+1}$, respectively. This results in at least $3l - 2$ path state changes for each time the shortcut is growing or shrinking, since the first and last state in each sequence do not incur a path state change. Since the shortcut switches at least $l$ times between growing and shrinking, we count at least $l \cdot (3l - 2) = 3l^2 - 2l = \Omega(n^2)$ path state changes during Phase III. □

We circumvent this issue by ignoring certain superfluous path state events.

Suppose that the shortcut is growing while the path $x$-$T$-$s_i$, for some $i = 1, 2, \ldots, k$, is a diametral path of type $x$-$T$-▲. In this situation a path state change may occur where the path $x$-$pq$-$\bar{x}$ or a path $x$-$pq$-$s_j$ for some $j > i$ might become a diametral path of type $x$-$pq$-● and $x$-$pq$-▲, respectively. There is no need to recognize this path state change, since the diameter cannot decrease before the shortcut becomes useful for $(x, s_i)$ again. Until then, we ignore any changes in the diametral path for pairs of type $x$-■. With this modification, the shortcut might leave the trajectory of the continuous algorithm. This does not compromise optimality, as we uphold the same invariants: the path $x$-$T$-$s_i$ certifies that no $x$-shift leads to a better shortcut, even if it is no longer diametral.

In Phase III, the path state has two components: the diametral path of type $x$-■ and the diametral path of type ■-$y$. We modify Phase III as follows. We ignore changes to the $x$-■-component when $x$-$T$-$s_i$ has become diametral for some $i \in \{1, 2, \ldots, k\}$ until $pq$ becomes useful for $(x, s_i)$, and we ignore changes to the ■-$y$-component when $s_j$-$T$-$y$ has become diametral for some $j \in \{1, 2, \ldots, k\}$ until $pq$ becomes useful for $(x, s_i)$. Figure 29 illustrates the modified Phase III for the geometric tree $T_l$ from the construction of the $\Omega(n^2)$ bound on the number of the path state changes from Lemma 16.

**Lemma 17** *For a geometric tree with n vertices and straight-line edges, the path state changes $O(n)$ times during the modified Phase III.*

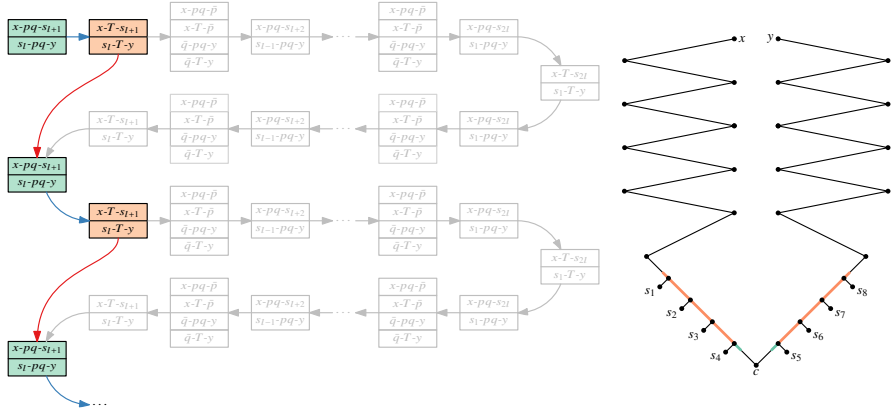

Figure 29: On the left, we compare the modified Phase III with the original Phase III for the tree $T_l$. Gray states and transitions are no longer visited by the modifed Phase III. Blue transitions occur while the shortcut is growing and red transitions occur while the shortcut is shrinking. On the right, we colour the sections of $T_l$ that are swept by $\bar{q}$ or $\bar{p}$ according to the corresponding path state. The number of path state changes decreased to $l-1$ with one state change for each time the shortcut changes between growing and shrinking.

PROOF. In Phase III, the speed of $p$ and $q$ is determined by one type of diametral path for a diametral pair of type $x$-■ and by one type of diametral path for a diametral pair of type ■-$y$, as indicated in Table 2. Certain path state transitions require the shortcut to shrink, others require it to grow. Suppose, for instance, that the shortcut shrinks as we move from $pq$ to $p'q'$, i.e., $|p'q'| < |pq|$. In this case, we cannot transit from $\{x\text{-}pq\text{-}\bullet, x\text{-}T\text{-}\bullet, \bullet\text{-}pq\text{-}y, \bullet\text{-}T\text{-}y\}$ to $\{x\text{-}pq\text{-}\blacktriangle, x\text{-}pq\text{-}\bullet, x\text{-}T\text{-}\bullet, \bullet\text{-}pq\text{-}y, \bullet\text{-}T\text{-}y\}$ as the shortcut moves from $pq$ to $p'q'$, since then $d_T(q,q') = d_T(p,p')$ and, thus, any path of type $x\text{-}pq\text{-}\blacktriangle$ shrinks by $d_{T+pq}(x, s_i) - d_{T+p'q'}(x, s_i) = d_T(p',p) + |pq| - |p'q'| - d_T(q',q) = |pq| - |p'q'|$ whereas the diameter only shrinks by $\frac{1}{2}(|pq| - |p'q'|)$. Figure 30 illustrates the path state transitions during Phase III for the path states that do not contain $x\text{-}T\text{-}\blacktriangle$ or $\blacktriangle\text{-}T\text{-}y$.

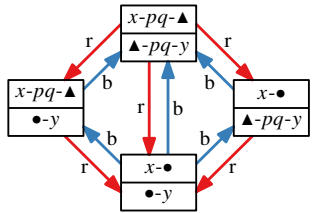

Figure 30: The path states encountered during Phase III, excluding the states containing $x\text{-}T\text{-}\blacktriangle$ or $\blacktriangle\text{-}T\text{-}y$ and any transitions to these states. Red transitions (r) occur while the shortcut is shrinking; blue transitions (b) occur while the shortcut is growing.

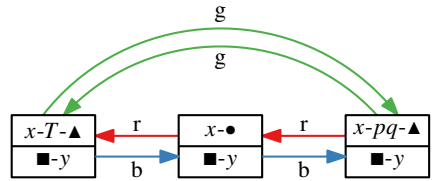

Figure 31: A simplified view on the path states encountered during Phase III. Only the changes in the path type for $x$-■ are shown. Red transitions (r) may occur while the shortcut is shrinking; blue transitions (b) may occur while the shortcut is growing.

Recall that the $\mathcal{B}$-sub-trees $S_1, S_2, \ldots, S_k$ are numbered in the order along the backbone from $a$ to $b$. After $x\text{-}T\text{-}s_i$ has become a diametral path of type $x\text{-}T\text{-}\blacktriangle$ in the modified Phase III, none of the paths $x\text{-}T\text{-}s_j$ with $i < j$ will be registered as diametral paths. If $x$-$T$-$s_j$ does become diametral after $x$-$T$-$s_i$ has become diametral, then the shortcut is useless for $(x, s_i)$, since it is useless for $(x, s_j)$. Therefore, we do not register that $x$-$T$-$s_j$ becomes diametral, because we are still waiting for the shortcut to become useful for $(x, s_i)$, since $d_T(x, s_i) = d_{T+pq}(x, s_i) \leq \text{diam}(T + pq)$.

We argue that the modified Phase III visits the path states containing $x$-$T$-▲ at most $k + n$ times. Figure 31 illustrates the possible changes in the $x$-■-component of the path state during the modified Phase III. When the shortcut grows, we only enter a path state containing $x$-$T$-▲ when the path $x$-$T$-$s_i$ that was most recently a diametral path of type $x$-$T$-▲ becomes diametral again. In this case, the $x$-■-component only changes away from $x$-$T$-▲ after the shortcut has began to shrink again. Since the shortcut is growing at most $n$ times during the modified Phase III, we register at most $O(n)$ path state changes of this kind. When the shortcut shrinks, we only enter a path state containing $x$-$T$-▲ when a path $x$-$T$-$s_l$ becomes diametral that has not been diametral before. As argued above, this may occur at most $k$ times, since $l < i$, where $x$-$T$-$s_i$ was the most recent diametral path of type $x$-$T$-▲. Likewise, we argue that the modified Phase III visits the path states containing ▲-$T$-$y$ at most $k + n$ times. Once we have exhausted the at most $2k + 2n$ visits to path states containing $x$-$T$-▲ or ▲-$T$-$y$, the remaining path state transitions form acyclic digraphs when the shortcut is shrinking and when the shortcut is growing, as depicted in Figure 30. Since the shortcut changes at most $2n$ times between shrinking and growing, we register at most $2k + 4n$ events where the path state changes. Therefore, we process $O(n)$ path state events throughout the modified Phase III. □

We detect path state events during Phase III in the same fashion as in Phase I and II, except for the following two differences. First, the detection of diametral paths of type $x$-$pq$-▲ and ▲-$pq$-$y$ changes, since $x$-$pq$-$y$ is no longer diametral. Second, we need to detect when a diametral path that connects a diametral pair of type ■-■ appears, since this marks the end of Phase III. This concerns diametral paths of type ▲-$T$-▲ and ▲-$pq$-▲, since we already detect ▲-$pq$-● and ▲-$T$-● when monitoring the candidates for diametral paths with an endpoint on the simple cycle $C(p, q)$ in $T + pq$.

$x$-$pq$-▲, ▲-$pq$-$y$ There are two cases in which a path $x$-$pq$-$s_i$, for $i \in \{1, 2, \ldots, k\}$ becomes a diametral path of type $x$-$pq$-▲ during the modified Phase III.

In the first case, the path $x$-$T$-$s_i$ is a diametral path of type $x$-$T$-▲ and the shortcut is about to become useful for $(x, s_i)$. We can detect this in constant time per vertex event by comparing $d_T(a, r_i)$ with $d_T(a, \bar{p})$ to see when and if $\bar{p}$ passes through $r_i$.

In the second case, the shortcut is growing and the path $x$-$pq$-$\bar{p}$ is a diametral path of type $x$-$pq$-●. The point $\bar{p}$ is moving towards $y$ when the shortcut is growing and $s_i$ is a leaf of a secondary $\mathcal{B}$-sub-tree attached to the path from $q$ to $\bar{p}$. This implies that we have $d_T(\bar{q}, r_i) = d_T(r_i, s_i)$ in this case. Therefore, we can detect this type of path state event by placing additional vertices at the points $\bar{p}_1, \bar{p}_2, \ldots, \bar{p}_k$ along the backbone such that $\bar{p}_i$ is the point along the path from $a$ to $r_i$ with $d_T(\bar{p}_i, r_i) = d_T(r_i, s_i)$, if such a point exists. Placing these at most $k$ additional vertices takes $O(n + k \log k)$ preprocessing time.

▲-*T*-▲ We ignore diametral paths of type ▲-*T*-▲. Suppose, during the modified Phase III, $s_i$-*T*-$s_j$, for $i, j \in \{1, 2, \ldots, k\}$, becomes a diametral path of type ▲-*T*-▲ for some shortcut position $\hat{p}\hat{q}$. Even if we fail to register this path state event, we still report an optimal shortcut, since we report the shortcut that yields the smallest encountered continuous diameter including the continuous diameter of $T + \hat{p}\hat{q}$.

▲-*pq*-▲ We cannot ignore an event where a diametral path of type ▲-*pq*-▲ appears, since the length of these paths depends on *pq*. Moreover, we cannot afford to check whether a diametral path of type ▲-*pq*-▲ is about to appear when processing the other events. We perform the detection of such events as a post-processing step instead. It is sufficient to find the first position $\hat{p}\hat{q}$ where some path $s_i$-*pq*-$s_j$, for $i, j \in \{1, 2, \ldots, k\}$, becomes a diametral path of type ▲-*pq*-▲: If we shift outwards from $\hat{p}\hat{q}$, then $s_i$-*pq*-$s_j$ will remain diametral while increasing in length. We proceed as follows. First, we simulate the modified Phase III without attempting to detect if a diametral path of type ▲-*pq*-▲ appears. We record the sequence of edge pairs that we visit during this simulation. As argued in Lemma 17, this sequence contains $O(n)$ edge pairs. After the simulation, we perform a binary search for $\hat{p}\hat{q}$ in the sequence of visited edge pairs. The binary search for $\hat{p}\hat{q}$ takes $O(n \log n)$ time, since we can determine the largest path of type ▲-*pq*-▲ in $O(n)$ time for a fixed position of the shortcut, as shown in Lemma 18.

**Lemma 18** *For every augmented tree $T + pq$ with $n$ vertices and straight-line edges, we can determine the length of the longest paths of type ▲-pq-▲ in $O(n)$ time.*

PROOF. Every path of type ▲-*pq*-▲ has length $d_T(s_i, p) + |pq| + d_T(q, s_j)$ for some $i, j = 1, 2, \ldots, k$ with $i < j$ such that $pq$ is useful for $(s_i, s_j)$. This means that $s_i$ is a leaf of one of the secondary $\mathcal{B}$-sub-tees $S_{i_L}, S_{i_L+1}, \ldots, S_{i_H}$ that are attached to the path from $p$ to $\bar{q}$, and $s_j$ is a leaf of one of the secondary $\mathcal{B}$-sub-tees $S_{j_L}, S_{j_L+1}, \ldots, S_{j_H}$ attached to the path from $\bar{p}$ to $q$. We prove that the matrix $M$ with entries

$$M_{j,i} = \begin{cases} d_T(s_i, p) + |pq| + d_T(q, s_j) & \text{, if } pq \text{ is useful for } (s_i, s_j) \\ 0 & \text{, otherwise} \end{cases} \quad \begin{array}{l} \text{for } i_L \leq i \leq i_H \\ \text{and } j_L \leq j \leq j_H \end{array},$$

is totally monotone. This means we have to show that $M_{j_1, i_1} < M_{j_1, i_2}$ implies $M_{j_2, i_1} < M_{j_2, i_2}$ for all indices $i_1, i_2, j_1, j_2$ with $i_L \leq i_1 < i_2 \leq i_R$ and $j_L \leq j_1 < j_2 \leq j_R$.

Suppose we have $M_{j_1, i_1} < M_{j_1, i_2}$. Then $M_{j_1, i_2} > 0$, because all entries of $M$ are non-negative. This means that $pq$ is useful for $(s_{i_2}, s_{j_1})$. Therefore, the shortcut $pq$ is also useful for $(s_{i_1}, s_{j_1})$, for $(s_{i_1}, s_{j_2})$, and for $(s_{i_2}, s_{j_2})$, due to the relative positions of $r_{i_1}, r_{i_2}, r_{j_1}$, and $r_{j_2}$, as shown in Figure 32. Hence, $M_{j,i} = d_T(s_i, p) + |pq| + d_T(q, s_j)$ for $i \in \{i_1, i_2\}$ and $j \in \{j_1, j_2\}$. With this observation, $M_{j_1, i_1} < M_{j_1, i_2}$ implies $d_T(s_{i_1}, p) < d_T(s_{i_2}, p)$, since $d_T(s_{i_1}, p) + |pq| + d_T(q, s_{j_1}) = M_{j_1, i_1} < M_{j_1, i_2} = d_T(s_{i_2}, p) + |pq| + d_T(q, s_{j_1})$. This implies $M_{j_2, i_1} < M_{j_2, i_2}$, because

$$M_{j_2, i_1} = d_T(s_{i_1}, p) + |pq| + d_T(q, s_{j_2}) < d_T(s_{i_2}, p) + |pq| + d_T(q, s_{j_2}) = M_{j_2, i_2} \ .$$

Thus, the matrix $M$ is totally monotone. We can access any entry $M_{j,i}$ of $M$ in constant time after $O(n)$ pre-processing: We determine $d_T(s_1, r_1), d_T(s_2, r_2), \ldots, d_T(s_k, r_k)$ as well as $d_T(a, r_1), d_T(a, r_2), \ldots, d_T(a, r_k)$ in advance. The shortcut $pq$ is useful for $(s_i, s_j)$ precisely when $d_T(r_i, p) + |pq| + d_T(q, r_j) < d_T(r_i, r_j)$, which we can check in constant time. Computing $d_T(s_i, p) + |pq| + d_T(q, s_j)$ also takes constant time. Therefore, we can determine a largest entry in $M$—and, thus, a longest path of type ▲-$pq$-▲—in $O(n)$ time using the SMAWK Algorithm [16] without constructing $M$ explicitly. □

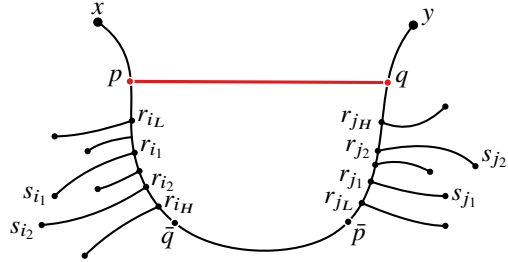

Figure 32: The relative positions of the secondary $\mathcal{B}$-sub-trees with indices $i_L, i_1, i_2, i_H, j_L, j_1, j_2,$ and $j_H$ when determining the longest ▲-$pq$-▲ path.

In conclusion, we can discretize all phases of the continuous algorithm—with some modifications that do not impact optimality—with $O(n)$ events that we can process in $O(n \log n)$ total time, followed by a post-processing step that takes $O(n \log n)$ time.

**Theorem 19** *For any geometric tree $T$ with $n$ straight-line edges, we determine a shortcut $pq$ that minimizes the continuous diameter of $T + pq$ in $O(n \log n)$ time.* □

## 7. Conclusion

We discussed the problem of minimizing the continuous diameter when augmenting a geometric tree with a single shortcut. A natural extension of this problem would be to minimize the continuous diameter when augmenting a geometric network with multiple shortcuts. For instance, given a number $k \geq 2$, we would like to characterize the trees where at least $k$ shortcuts are required to reduce the continuous diameter.

While our structural results hold for geometric trees whose edges are arbitrary rectifiable curves, the algorithmic results are described for geometric trees with straight-line edges. However, the algorithmic results extend to more general types of edges, as well. For instance, if the edges are algebraic curves of degree $d$, then the running time of the algorithm becomes $O(dn \log n)$, since the shortcut may switch $O(d)$ times between growing and shrinking when both endpoints slide along a fixed pair of edges. We leave it to future research to identify the minimum requirements for our algorithmic results.